\documentclass[acmsmall,dvipsnames,screen]{acmart}
\setcopyright{rightsretained}
\acmJournal{PACMPL}
\acmYear{2024} 
\acmVolume{8} 
\acmNumber{OOPSLA2} 
\acmArticle{335} 
\acmMonth{10}
\acmDOI{10.1145/3689775}

\usepackage{booktabs}   
\usepackage{subcaption} 

\setcitestyle{round}

\usepackage{paralist}
\usepackage{listing}
\usepackage[ruled,algosection,noend]{algorithm2e}
\usepackage{multirow}
\usepackage{prooftree}
\usepackage{varwidth}
\usepackage{graphicx}
\usepackage{listings}
\usepackage{float}
\usepackage{multirow}
\usepackage[noend]{algorithmic}
\usepackage{mathpartir}
\usepackage{url}
\usepackage{hyperref}
\usepackage[capitalise]{cleveref}
\usepackage{xspace}
\usepackage{verbatim}
\usepackage{lipsum}
\usepackage{wrapfig}
\usepackage{ifsym}

\usepackage[inline, shortlabels]{enumitem}
\usepackage{url}
\usepackage{fancyvrb}
\usepackage[figurename=Figure]{caption}
\usepackage{nanoc}
\usepackage[T1]{fontenc}
\usepackage{relsize}
\usepackage{marvosym}
\usepackage{microtype}
\usepackage[frozencache,cachedir=.]{minted}

\usepackage{adjustbox}
\usepackage{multicol}
\usepackage{xcolor}
\usepackage{tikz}
\usepackage{tcolorbox}
\usepackage{ragged2e}
\usepackage{pgfplots}
\usepackage{pgfplotstable}

\usetikzlibrary{arrows}
\usetikzlibrary{shapes.arrows}
\usetikzlibrary{shapes}
\usetikzlibrary{calc}
\usetikzlibrary{positioning}
\usetikzlibrary{fit}
\usetikzlibrary{tikzmark}
\usetikzlibrary{snakes}

\usepgfplotslibrary{groupplots}

\usepackage{tikzsymbols} 

\tikzstyle{arrow}+=[thick,rounded corners=0.5em]
\tikzstyle{every picture}+=[remember picture,baseline]

\usepackage{multirow}
\usepackage{hhline}

\hypersetup{colorlinks,
  linkcolor=ACMDarkBlue,
  citecolor=ACMPurple,
  urlcolor=ACMDarkBlue,
  filecolor=ACMDarkBlue}

\floatname{algorithm}{Algorithm}

\SetAlFnt{\small}
\SetAlCapFnt{\small}
\SetAlCapNameFnt{\small}
\SetAlCapHSkip{0pt}
\IncMargin{-\parindent}

\floatstyle{plaintop}
\restylefloat{table}

\defcitealias{Coq-manual}{Coq}

\makeatletter 
\def\arcr{\@arraycr}
\makeatother

\definecolor{shadecolor}{gray}{1.00}
\definecolor{darkgray}{gray}{0.30}
\definecolor{violet}{rgb}{0.56, 0.0, 1.0}
\definecolor{forestgreen}{rgb}{0.13, 0.55, 0.13}
\definecolor{mygray}{rgb}{0.5,0.5,0.5}

\definecolor[named]{ACMBlue}{cmyk}{1,0.1,0,0.1}
\definecolor[named]{ACMYellow}{cmyk}{0,0.16,1,0}
\definecolor[named]{ACMOrange}{cmyk}{0,0.42,1,0.01}
\definecolor[named]{ACMRed}{cmyk}{0,0.90,0.86,0}
\definecolor[named]{ACMLightBlue}{cmyk}{0.49,0.01,0,0}
\definecolor[named]{ACMGreen}{cmyk}{0.20,0,1,0.19}
\definecolor[named]{ACMPurple}{cmyk}{0.55,1,0,0.15}
\definecolor[named]{ACMDarkBlue}{cmyk}{1,0.58,0,0.21}
\definecolor[named]{PaleGreen}{RGB}{196, 255, 231}
\definecolor[named]{PaleOrange}{RGB}{255, 213, 169}
\definecolor{intnull}{RGB}{213,229,255}

\definecolor{shadecolor}{gray}{1.00}
\definecolor{ddarkgray}{gray}{0.85}
\definecolor{darkgray}{gray}{0.30}
\definecolor{light-gray}{gray}{0.91}
\definecolor{lgray}{gray}{0.96}

\newcommand{\etc}{\emph{etc}\xspace}
\newcommand{\ie}{\emph{i.e.}\xspace}

\newcommand{\eg}{\emph{e.g.}\xspace}

\newcommand{\aka}{\textit{a.k.a.}\xspace}
\newcommand{\cf}{\textit{cf.}\xspace}

\newcommand{\tname}[1]{\textsc{#1}\xspace}
\newcommand{\domainslib}{\textsf{Domainslib}\xspace}

\newcommand{\tool}{\tname{OBatcher}}

\definecolor{pblue}{rgb}{0.13,0.13,1}
\definecolor{pgreen}{rgb}{0,0.5,0}
\definecolor{pred}{rgb}{0.9,0,0}
\definecolor{pgrey}{rgb}{0.46,0.45,0.48}

\definecolor{ckeyword}{HTML}{7F0055}
\definecolor{ccomment}{HTML}{3F7F5F}
\definecolor{cnumber}{HTML}{2A0099}

\setmintedinline{fontsize=\small}
\newcommand{\ocode}[1]{\mintinline[fontsize=\small]{ocaml}{#1}}
\newcommand{\code}[1]{\ocode{#1}}

\newcommand{\set}[1]{\left\{{#1}\right\}}

\makeatletter
\protected\def\ccell#1#{%
  \kern-\fboxsep
  \@ccell{#1}%
}
\def\@ccell#1#2#3{%
  \colorbox#1{#2}{#3}%
  \kern-\fboxsep
}
\makeatother

\newcommand{\pts}{\mapsto}

\newcommand{\sep}{\ast}

\newcommand{\hide}[1]{}

\makeatletter
\providecommand*{\cupdot}{%
  \mathbin{%
    \mathpalette\@cupdot{}%
  }%
}
\newcommand*{\@cupdot}[2]{%
  \ooalign{%
    $\m@th#1\cup$\cr
    \hidewidth$\m@th#1\cdot$\hidewidth
  }%
}
\makeatother

\DeclareMathSymbol{\synth}{\mathrel}{symbolsC}{123}

\tikzset{
  proof/.pic={
      \draw[pic actions] (-0.25,-0.25) .. controls ++(0.125,0.25) and ++(-0.125,-0.25) .. (-0.25,0.25)
             -- (0.25,0.25) .. controls ++(-0.125, -0.25) and ++(0.125, 0.25) .. (0.25,-0.25) 
             -- cycle;
             \node at (0, 0) {#1};
  }
}

\tikzstyle{proof}=[thin, fill=blue!20, minimum width=1em]
\tikzstyle{component}=[thin, fill=ForestGreen!20, draw=none, align=center, rounded corners=0pt]
\tikzstyle{generator}=[thin, fill=OrangeRed!20, draw=none, align=center, rounded corners=1pt]
\tikzstyle{validator}=[thin, fill=Orchid!20, draw=none, align=center, rounded corners=0pt]
\tikzstyle{instantiator}=[thin, fill=Red!15, draw=none, align=center, rounded corners=0pt]
\tikzstyle{proof step}=[>=stealth,->,semithick]

\definecolor{shadecolor}{gray}{1.00}
\definecolor{darkgray}{gray}{0.30}
\definecolor{violet}{rgb}{0.56, 0.0, 1.0}
\definecolor{forestgreen}{rgb}{0.13, 0.55, 0.13}
\definecolor{mygray}{rgb}{0.5,0.5,0.5}
\definecolor{mahogany}{RGB}{192, 64, 0}

\lstdefinelanguage{Coq} {
mathescape=true,						
texcl=false,
morekeywords=[1]{
  Add,
  All,
  Arguments,
  Axiom,
  Bind,
  Canonical,
  Check,
  Close,
  CoFixpoint,
  CoInductive,
  Coercion,
  Contextual,
  Corollary,
  Defined,
  Definition,
  Delimit,
  End,
  Example,
  Export,
  Fact,
  Fixpoint,
  Goal,
  Graph,
  Hint,
  Hypotheses,
  Hypothesis,
  Implicit,
  Implicits,
  Import,
  Inductive,
  Lemma,
  Let,
  Local,
  Locate,
  Ltac,
  Maximal
  Module,
  Morphism,
  Next,
  Notation,
  Obligation,
  Open,
  Parameter,
  Parameters,
  Prenex,
  Print,
  Printing,
  Program,
  Projections,
  Proof,
  Proposition,
  Qed,
  Record,
  Relation,
  Remark,
  Require,
  Reserved,
  Resolve,
  Rewrite,
  Save,
  Scope,
  Search,
  Section,
  Show,
  Strict,
  Structure,
  Tactic,
  Theorem,
  Unset,
  Variable,
  Variables,
  View,
  inside,
  outside
},
morekeywords=[2]{
  as,
  cofix,
  false,
  else,
  end,
  exists,
  exists2,
  fix,
  for,
  forall,
  fun,
  if,
  in,
  is,
  let,
  match,
  nosimpl,
  of,
  return,
  struct,
  then,
  true,
  vfun,
  with
},
morekeywords=[3]{Type, Prop, Set, True, False},
morekeywords=[4]{
  ptr, seq, nat, unit, heap,
  after,
  apply,
  assert,
  auto,
  bool_congr,
  case,
  change,
  clear,
  compute,
  congr,
  constructor,
  cut,
  cutrewrite,
  destruct,
  elim,
  field,
  fold,
  generalize,
  have,
  hhauto,
  heval, 
  hnf,
  induction,
  injection,
  intro,
  intros,
  intuition,
  inversion,
  left,
  loss,
  move,
  nat_congr,
  nat_norm,
  pattern,
  pose,
  refine,
  rename,
  replace,
  revert,
  rewrite,
  right,
  ring,
  set,
  simpl,
  split,
  suff,
  suffices,
  symmetry,
  transitivity,
  trivial,
  unfold,
  unlock,
  using,
  without,
  wlog,
  autorewrite
},        
morekeywords=[5]{
  assumption,
  by,
  contradiction,
  done,
  exact,
  lia,
  gappa,
  omega,
  reflexivity,
  romega,
  solve,
  tauto,
  discriminate,
  unsat
},
morecomment=[s]{(*}{*)},
morekeywords=[6]{do, first, try, idtac},
showstringspaces=false,
morestring=[b]",
tabsize=3,							
extendedchars=true,  		 		
sensitive=true, 
breaklines=false,
basicstyle=\footnotesize\ttfamily,
captionpos=b,							
columns=[l]fullflexible,
identifierstyle={\color{black}},
keywordstyle=[1]{\color{violet}},
keywordstyle=[2]{\color{forestgreen}\bfseries},
keywordstyle=[3]{\color{forestgreen}\bfseries},
keywordstyle=[4]{\color{blue}},
keywordstyle=[5]{\color{red}},
keywordstyle=[6]{\color{violet}},
stringstyle=\it\ttfamily\color{mahogany},
commentstyle=\it\ttfamily\color{Bittersweet},
numberstyle=\tiny\color{mygray},
literate={\\/}{{$\vee$}}1
         {/\\}{{$\wedge$~}}1
         {:->}{{$\mapsto~$\!}}1
         {ent}{{$\vdash~$}}1
         {\\in}{{$\in~$}}1
         {++}{{$+\!+\!~$}}1
         {.+}{{$\!\!+$}}1
         {-->}{{$\rightarrow$}}1
         {forall}{{$\forall~$}}1
         {not}{{$\neg~$}}1
         {exists}{{$\exists~$}}1
         {sep}{{$\sep~$}}1
         {\\+}{{$\!\join\!~$}}1
         {capsigma}{{$\Sigma~$}}1
         {lceil}{{$\lceil~$}}1
         {rceil}{{$\rceil~$}}1
         {alpha}{{$\alpha~$}}1
         {beta}{{$\beta~$}}1
}

\lstdefinestyle{Coq}{language=Coq}

\setlist[itemize]{leftmargin=*}
\setlist[enumerate]{leftmargin=*}

\allowdisplaybreaks

\begin{document}

\newcommand{\mytitle}{Concurrent Data Structures Made Easy}

\newcommand{\runningtitle}{\mytitle}

\setlength\floatsep{1.25\baselineskip plus 3pt minus 2pt}
\setlength\textfloatsep{1.25\baselineskip plus 3pt minus 2pt}
\setlength\intextsep{1.25\baselineskip plus 3pt minus 2 pt}

\title[\mytitle]{\mytitle}
\subtitle{Extended Version}

\author{Callista Le}
\affiliation{%
  \institution{Yale-NUS College}
    \country{Singapore}
}
\email{phongnguyen.le@u.yale-nus.edu.sg}
\orcid{0009-0009-6689-9542}

\author{Kiran Gopinathan}
\affiliation{%
  \institution{National University of Singapore}
    \country{Singapore}
}
\email{mail@kirancodes.me}
\orcid{0000-0002-1877-9871}

\author{Koon Wen Lee}
\affiliation{%
  \institution{Ahrefs}
    \country{Singapore}
}
\email{koonwen@gmail.com}
\orcid{0009-0008-0971-7530}

\author{Seth Gilbert}
\affiliation{%
  \institution{National University of Singapore}
    \country{Singapore}
}
\email{seth.gilbert@comp.nus.edu.sg}
\orcid{0000-0003-3298-7412}

\author{Ilya Sergey}
\affiliation{%
  \institution{National University of Singapore}
    \country{Singapore}
}
\email{ilya@nus.edu.sg}
\orcid{0000-0003-4250-5392}

\begin{abstract}
%
%
Design of an efficient thread-safe concurrent data structure is a
balancing act between its implementation complexity and performance.
%
%
\emph{Lock-based} concurrent data structures, which are relatively
easy to derive from their sequential counterparts and to prove
thread-safe, suffer from poor throughput under even light
multi-threaded workload.
At the same time, \emph{lock-free} concurrent structures allow for
high throughput, but are notoriously difficult to get right and
require careful reasoning to formally establish their correctness.


%
In this work, we explore a solution to this conundrum based on a
relatively old idea of \emph{batch parallelism}---an approach for
designing high-throughput concurrent data structures via a simple
insight: efficiently processing a batch of \emph{a priori known}
operations in parallel is easier than optimising performance for a
stream of arbitrary asynchronous requests.
Alas, batch-parallel structures have not seen wide practical adoption
due to ($i$)~the inconvenience of having to structure multi-threaded
programs to explicitly group operations and ($ii$)~the lack of a
systematic methodology to implement batch-parallel structures as
simply as lock-based ones.

We present \tool---a Multicore OCaml library that streamlines the
design, implementation, and usage of batch-parallel structures.
It solves the first challenge (how to use) by suggesting a new
lightweight \emph{implicit batching} design that is built on top of
generic asynchronous programming mechanisms.
The second challenge (how to implement) is addressed by identifying
\emph{a family of strategies} for converting common sequential
structures into efficient batch-parallel ones,
and by providing functors that embody those strategies.
We showcase \tool with a diverse set of benchmarks. 
%
%
Our evaluation of all the implementations on large asynchronous
workloads shows that (a)~they consistently outperform the
corresponding coarse-grained lock-based implementations and that
(b)~their throughput scales reasonably with the number of processors.

\vspace{10pt}
\end{abstract}




\maketitle

\vspace{5pt}

\section{Introduction}
\label{sec:intro}

Mutable concurrent data structures are a key component of
multi-threaded programs that run on common shared-memory machines.
Efficient concurrent data structures, which allow for high degree of
parallelism, are difficult to design and even more difficult to prove
correct due to the intricate invariants that have to be to be
preserved by the possibly overlapping concurrent operations.

\newpage


Given a particular sequential data structure, converting it into a
thread-safe concurrent counterpart is usually done by following one of
the two approaches.
In the first approach, all operations of the data structure are made
\emph{synchronised}, \ie, protected by a single lock that must be
acquired by a thread before invoking any of them and released
afterwards.
The main advantage of this approach, known as \emph{coarse-grained}
concurrency, is that it is easy to implement, making the migration of
sequential code to concurrent almost mechanical.
The main downside is that it removes any opportunity for parallelism
\emph{within} the data structure itself, effectively making all
operations with it mutually-exclusive, and, therefore, introducing
sequential bottlenecks.

The second approach, dubbed \emph{fine-grained} concurrency, requires
one to carefully consider possible interactions between multiple
concurrent operations of a data structure that can overlap in time,
while being executed by different threads, introducing synchronisation
sparingly to ensure the correctness of the updates and query results.
The reduced amount of synchronisation allows multiple threads to
execute the data structure's operations in parallel, improving its
throughput at the cost of significantly increased conceptual
complexity: fine-grained concurrent data structures are known to be
extremely difficult to design and implement correctly, requiring
non-trivial expertise~\cite{AoMP}, and their formal verification is
still an active research field to
date~\cite{SergeyNB15,FeldmanKE0NRS20,MulderKG22,MeyerWW22}.

The technique of \emph{flat combining} by~\citet{HendlerIST10} was an
initial attempt to bridge the gap between coarse and fine-grained
concurrency. It presented a lock-sharing discipline for coarse-grained
data structures that amortised the cost of repeatedly locking and
unlocking the structure by instead promoting an arbitrary client to
temporarily serve as a dedicated worker that would lock the data
structure once and sequentially handle several client requests in a
batch.
While this strategy was effective at reducing lock contentention and
thus the overheads of locking, flat combining was not concerned with
exploiting parallelism in the data structure itself, leaving the
entire construction still substantially worse than a dedicated
fine-grained implementation.

\emph{Batch-parallel data structures} have been
proposed as an alternative design pattern that aims to offer a
trade-off between parallel performance and the implementation
complexity.
Batch-parallel data structures differ from traditional concurrent data
structures in that they process a \emph{batch} of operations
collectively in parallel, instead of handling arbitrary asynchronously
incoming requests one by one.
Within the batch, such structures exploit parallelism by dynamically
spawning asynchronous
computations~\cite{DhulipalaDKPSS20,AcarABDW20,DriscollGST88,PaigeK85,BrodalTZ98}.
This design comes with multiple advantages:

\begin{enumerate}
\item \label{b1} parallel processing of a batch can be optimised
  based on the properties of its operations;
\item \label{b2} the structure can control the order in which the
  operations in a batch are processed;
\item \label{b3} batch-parallel data structures can be systematically
  derived from their sequential versions and incrementally retrofitted
  with more parallelism.
\end{enumerate}

The main reason why batch-parallel data structures have not been taken
up by practitioners is because structuring programs to pass an
\emph{explicit batch} of operations is inconvenient and unfeasible in
common asynchronous contexts.
\emph{Implicit batching} by~\citet{AgrawalFSSU14} is a technique meant
to circumvent this problem by means of providing a custom scheduler to
transparently batch individual requests made by client threads before
they are sent to a batch-parallel data structure.
The only existing prototype implementation of implicit batching has
been done by its authors in Cilk-5~\cite{FrigoLR98} by explicitly
modifying its runtime scheduler's internals---hardly a lightweight
solution that could be easily reproduced in most modern programming
languages.
Furthermore, \citeauthor{AgrawalFSSU14}'s implementation of implicit
batching does not allow to have \emph{more than one} batch-parallel
data structure per application, which renders it impractical for
real-world tasks.

Given the state of the art, we phrase the motivation for this work as
the following question:
\begin{center}
{
  \begin{minipage}{\textwidth-12em}
\[
    \begin{array}{c}
    \text{Can we implement \emph{efficient}, \emph{easy-to-understand}, and
    \emph{easy-to-use}}
    \\
    \text{concurrent data structures via batch parallelism?}  
    \end{array}
\]
  \end{minipage}
}
\end{center}

In the rest of this paper, we answer this question affirmatively.


\paragraph{Key ideas}

We observe that the technique of implicit batching can be implemented
in a lightweight way \emph{as a library} by re-purposing techniques
from flat combining in conjunction with common concurrent programming
primitives, namely \code{async}/\code{await}, which are available in
many modern mainstream programming languages, including Python, Rust,
Kotlin, Swift, and, since recently, OCaml.
In particular, in OCaml 5 (\aka Multicore OCaml), high-level libraries
such as \textsf{Eio} and \domainslib implement the primitives for
asynchronous programming using the mechanism of \emph{effect
  handlers}~\cite{Sivaramakrishnan21}.
It turns out that the support for asynchronous programming, as
provided by \domainslib, is sufficient to address the two
shortcomings of \citeauthor{AgrawalFSSU14}'s proposal and allows one
to (1)~implement implicit batch parallelism without modifying the
runtime scheduler, and (2)~accommodate multiple batch-parallel
concurrent data structures within a single program, fairly
distributing available computational resources between their
operations.

Having attempted to implement batch-parallel versions of common
sequential search structures, such as AVL and van Emde Boas trees, we
have noticed that many of them fall into one of the two categories
when it comes to orchestrating their internal parallelism.
In particular, operations of tree-like search structures, whose
sub-structures (\ie, subtrees) are themselves valid instances of that
data structure, admit a principled parallelisation strategy based on
the \emph{split-join} idea by~\citet{BlellochFS16} and the concept of
bulk updates \cite{Akhremtsev2016,Sanders2019}.
In contrast, for search structures that rely on prefixes or hashes to
store keys and do not admit a natural ``splitting'', parallel
execution of operations can be done by exploiting the ``locality'' of
their effect---a novel idea that we call \emph{expose-repair}.
We capture these observations, as well as a generic structure for
batch-parallelism, in an extensible higher-order OCaml library that
features functors for the batch parallelism strategies described above
and makes it straightforward to implement batch-parallel versions for
many common sequential data structures.

\paragraph{Contributions}

To summarise, in this work we make the following contributions.

\begin{itemize}
\item Our main practical contribution is \tool: a Multicore OCaml
  library that facilitates implementation of batch-parallel
  structures, while making their usage transparent to the
  clients~(\autoref{sec:design}).
  \tool is \emph{lightweight}: it does not require any significant
  changes in the client code to use it, and implementations in it can
  be used as drop-in replacements for their coarse-grained analogues.
  As a framework, \tool allows one to develop new batch-parallel
  structures \emph{gradually} by elaborating the implementation of a
  function that processes a batch of operations---with the default
  implementation simply doing so sequentially, with no parallelism
  whatsoever.

\item Our main conceptual contribution is the observation that many
  search structures, whose concurrent versions are traditionally
  considered challenging to implement, fall into one of the two
  categories, so-called ``split-join'' or ``expose-repair'', that
  allow for simple and efficient batch parallelism.
  We substantiate this observation by providing a family of OCaml
  functors built on top of \tool that abstract away the details of
  those batching strategies, streamlining implementation of concurrent
  versions of the respective sequential data structures
  (\autoref{sec:examples}).
  %
  %

\item We showcase \tool by implementing in it a wide range of
  concurrent search structures, including AVL, Red-Black, and van Emde
  Boas trees, treaps, x-fast and y-fast tries, B-trees, skip lists,
  and a concurrent version of a third-party Datalog solver.
  Our implementations outperform their coarse-grained counterparts in
  nearly all scenarios under diverse workloads
  (\autoref{sec:evaluation}).
\end{itemize}

\noindent
Though we use OCaml as the primary language for our implementation and
presentation due to its convenient mechanisms for composition via
functors and for concurrency via \domainslib's primitives, the core
ideas of this work generalise beyond a single language and can be
easily replicated in any modern language that supports
\code{async}/\code{await}-style concurrency with multi-producer,
multi-consumer channels and task pools, as well as means for code
reuse via mechanisms such as type classes or traits. We substantiate
this claim with a Rust implementation of the core framework.

\section{Overview}
\label{sec:overview}


We start by building an intuition for \tool with a help of a simple
concurrent application in OCaml that keeps track of the number of
visiting users in a multi-core web-server.

To begin, consider a request handler that outputs the total number of
seen visitors of the server:
%
%
\begin{minted}[fontsize=\footnotesize,
% linenos,numbersep=7pt,xleftmargin=15pt,
bgcolor=lgray]{ocaml}
let c = Counter.init ()
let handle_request = fun _ -> 
    Counter.incr c; 
    printf "you are the %d'th visitor!" (Counter.get c)
\end{minted}
In this program, each time a request is received, the handler calls
out to a \code{Counter} module, first incrementing the count \code{c}
of seen visitors, and then retrieving the current value to print out a
welcome message.
While this handler could easily be ported from a vanilla OCaml
codebase to a multi-core one, the same cannot be said for the
underlying \code{Counter} module that it relies upon.


\subsection{A Straw-Man Solution: Coarse-Grained Locking}
\label{sec:coarse-grained}
Classically, OCaml programs have had the benefit of assuming a
single-threaded execution, so a vanilla implementation of
\code{Counter} could simply be a wrapper around a
reference (\cf~\autoref{fig:vanilla-counter}).

\begin{figure}[t]
  \begin{subfigure}{0.45\textwidth}
  \begin{minipage}{1.0\textwidth}
\begin{minted}[fontsize=\footnotesize,linenos,numbersep=7pt,xleftmargin=15pt,bgcolor=lgray]{ocaml}
module Counter = struct
  type t = int ref
  let init () = ref 0
  let incr c =
      c := !c + 1
  let get c = !c

end
      \end{minted}
  \end{minipage}
  \caption{A sequential counter}
  \label{fig:vanilla-counter}
  \end{subfigure}
  \begin{subfigure}{0.53\textwidth}
  \begin{minipage}{1.0\textwidth}
\begin{minted}[fontsize=\footnotesize,linenos,numbersep=7pt,xleftmargin=15pt,bgcolor=lgray]{ocaml}
module CoarseCounter = struct
  type t = int ref * Mutex.t
  let init ()     = ref 0, Mutex.make ()
  let incr (c, l) = Mutex.with_lock l
                      (fun () -> c := !c + 1)
  let get  (c, l) = Mutex.with_lock l
                      (fun () -> !c)
end
    \end{minted}
   \end{minipage}
  \caption{A coarse-grained counter}
  \label{fig:coarse-grained-counter}
  \end{subfigure}
\caption{A sequential and a coarse-grained lock-based counter.}
\label{fig:counters}
\end{figure}

In the multi-core setting, these operations would not be atomic and
introduce a data race in the presence of concurrent calls to
\code{incr} and \code{get} of the same counter instance.
To avoid this, the operations of \code{Counter} must be written such
that they remain correct even in the presence of concurrent
executions, for example, by adopting so-called \emph{coarse-grained}
locking (\autoref{fig:coarse-grained-counter}).
This would make all manipulations with the counter
\emph{mutually-exclusive}, so any concurrent calls to them from
different threads would take place sequentially, removing any
opportunity for parallelism.

Using coarse-grained locking serves as a general mechanism for
migrating existing OCaml code to multi-core, but it leaves an
unsatisfying conclusion: we must explicitly rule out any concurrent
operations on the underlying data structure.
%
%
Conversely, constructing an efficient concurrent thread-safe
implementation requires careful analysis of how these operations
interact, and likely requires entirely redesigning the original
structure to introduce \emph{fine-grained} (\ie, internal)
synchronisation. 
Of course, for this pedagogical example, it would be simple to rewrite
it to use an atomic variable, but this can not be said of most data
structures that might need be ported to a concurrent setting.

Is it possible to find a middle ground between these two extremes, and
allow developers to utilise parallelism without having to fully give
up on their sequential implementations? With \tool, we answer this
question in the affirmative, using the idea of operation
\emph{batching} to bridge the gap.


\subsection{A Batched Interface for Counter}
\label{sec:ebatcher}

How can we exploit parallelism in the implementation of
\code{Counter}?
Unfortunately our current interface where operations are processed
individually does not leave much room for optimisation, so
let us now consider a different architecture, inspired by
\citeauthor{HendlerIST10}'s flat combiner.

Suppose our interface to \code{Counter} instead took in \emph{batches} of
operations to be executed in parallel:
\begin{center}
  \begin{minipage}{0.6\linewidth}
\begin{minted}[fontsize=\small]{ocaml}
type op = Get of int -> unit | Incr of unit -> unit
val run_batch: t -> op array -> unit
\end{minted}
\end{minipage}
\end{center}
Here, each reified operation \code{op} includes a callback to allow
the return value to be communicated to the caller.
To avoid excessive type parameterisations, all callbacks in \code{op}
have a return type \code{unit}, with the intuition that they are
effectful functions---as is the case, \eg, with our counting server.

The signatures above describe a new batched interface for the data
structure, which now admits a parallel implementation as presented in
\autoref{fig:runbatch}.
This new implementation uses the helper function
\code{parallel_reduce} to perform a map-reduce style parallel
computation. Internally, \code{parallel_reduce} partitions the
operation array into chunks, which are then each mapped to integer
deltas in parallel and then summed together to produce the total delta
to be applied to the counter.
As each submitted operation is processed, the mapping function also
invokes callback \code{kont} with an appropriate return value,
informing the client of the corresponding operation's completion.

\begin{wrapfigure}[10]{r}{0.35\linewidth}
\vspace{-15pt}
\setlength{\abovecaptionskip}{5pt}
\begin{minipage}{1.0\linewidth}
\begin{minted}[fontsize=\footnotesize,
%
% linenos,numbersep=0pt, xleftmargin=7pt,
%
bgcolor=lgray]{ocaml}
let run_batch counter batch =
  let vl : int = !counter in
  let delta = parallel_reduce
    (function
     | Get kont  -> kont vl; 0
     | Incr kont -> kont (); 1
     ) (+) batch in
  counter := !counter + delta
\end{minted}
\end{minipage}
\caption{Batch executor}
\label{fig:runbatch}
\end{wrapfigure}
The careful reader may be wondering about the correctness of this
implementation of \code{run_batch}.
In particular, observe that operations may sometimes produce seemingly
stale values as their results, such as the value \code{vl} passed to
\code{Get}'s callback.
Fortunately, this behaviour is correct, as it is justified from the
perspective of \emph{linearisability}~\cite{HerlihyW90}---a commonly
adopted correctness criterion for concurrent data structures.
In short, such ``stale'' effects of the \code{Get} operations from a
batch in the presence of concurrent interference correspond to a
collection of concurrent calls to the \code{get} method, which happen
to take effect \emph{right before} any increment operations taking place at the same time.\footnote{Or, in the concurrent programming jargon, these
  calls \emph{get linearised} before any concurrent increments.}
%

Overall, this new implementation provides increased throughput as
compared to the coarse-grained wrapper, but comes at the cost of
placing more restrictions on the caller, making it challenging to
integrate into existing code.
Firstly, in contrast to our previous direct-style interface where
requests are processed immediately, this new batched API expects a
\emph{collection} of requests to be submitted all at once, using
callbacks to return their results.
Furthermore, from a practical standpoint, the \code{kont} callbacks
should not be continuations of the client's code, as that would return
control from the batcher and the thread would be busy computing the
client's logic rather than processing other operations in the batch.
As it turns out, it is possible to elegantly solve these two problems
using the interface provided by Multicore OCaml and \domainslib.


\subsection{Implicit Batching via Asynchronous Programming}
\label{sec:async}

%

Let us solve the two shortcomings of the explicitly-batched counter
implementation described in \autoref{sec:ebatcher}: (1)~the need to
explicitly provide a collection of operations and the continuation and
(2)~the blocking invocation of the user's continuation \code{kont}
within the code \code{run_batch} in \autoref{fig:runbatch}.

\begin{wrapfigure}[11]{r}{0.48\textwidth}
\vspace{-3pt}
\setlength{\belowcaptionskip}{-15pt}
\setlength{\abovecaptionskip}{5pt}
\begin{minipage}{1.0\linewidth}
\begin{minted}[fontsize=\footnotesize,
%
linenos,numbersep=3pt, xleftmargin=11pt,
%
bgcolor=lgray]{ocaml}
let incr t (* counter & channel *) : unit =
  let promise: unit promise, 
      resolve: unit -> unit =
    Task.promise () in 
  Chan.send t.chan (Incr resolve);
  if (* no currently running batch *) then 
    let batch : op array = 
          Chan.collect_all t.chan in
    run_batch t.data pool batch;
  await promise
\end{minted}
\end{minipage}
\caption{Batched increment with a promise}
\label{fig:incr}
\end{wrapfigure}
The core design of \tool's construction is presented in
\autoref{fig:incr}.
When a client calls an operation on the batched data structure in a
direct style (\eg, incrementing the counter via \code{incr}), 
\tool first uses \domainslib to allocate a promise--a concurrent cell
primitive that can be used to store a result asynchronously--using
\code{Task.promise} (lines 2-4); this call returns the promise itself
and a function that can be called to resolve its value at a later
point.
The code then enqueues an \code{Incr} object on a channel associated
to this instance, \code{t.chan}, passing in the resolution function as
an argument.
\tool then checks whether a batch is currently running, and if not
promotes the current client to immediately execute the batch of
accummulated operations for the data structure (lines~6-9). Finally,
\tool completes the direct-style interface using the control inversion
provided by \domainslib mechanisms to \emph{suspend} the current
client task, via the \code{await} operation (line 10).
Intuitively, by awaiting on this newly allocated promise, the
continuation of the \code{incr} operation caller is effectively
suspended until the \code{resolve} function passed to \code{Incr} is
called.
This indirection through the promise object means that implementation
of \domainslib task scheduling will \emph{not} invoke the
continuation directly where it is called (as is done, \eg, in
\autoref{fig:runbatch}) but will instead reschedule its resumption as
a \emph{separate task} in a shared thread pool, thefore, avoiding the
blocking issue with the \code{run_batch} implementation.


\subsection{Putting It All Together}
\label{sec:together}

Let us now return back to our original example, and see how we can
instantiate \tool to use an explicitly-batched \code{Counter} with our
web server's request handler.

\begin{figure}[b]
  \centering
  \begin{subfigure}{0.45\textwidth}
  \begin{minipage}{1.0\textwidth}
  \begin{minted}[fontsize=\footnotesize,bgcolor=lgray]{ocaml}
module type Batched = sig
  type t
  type 'a op

  type wr_op = 
     Mk : ('a op * ('a -> unit)) -> wr_op
  val init: unit -> t
  val run_batch: 
     t -> pool -> wr_op array -> unit
end
\end{minted}
\end{minipage}
\caption{An explicitly-batched interface}
\label{fig:explicitly-batched-interface}
\end{subfigure}
  \begin{subfigure}{0.45\textwidth}
  \begin{minipage}{1.0\textwidth}
\begin{minted}[fontsize=\footnotesize,bgcolor=lgray]{ocaml}
module BatchedCounter : Batched = struct
  type t = int ref
  type 'a op = Get: int op 
             | Incr: unit op
  type wr_op = 
     Mk : ('a op * ('a -> unit)) -> wr_op
  let init () = ref 0
  let run_batch counter pool batch = 
       (* use parallel sum *)
end
\end{minted}
\end{minipage}
\caption{An example instantiation}
\label{fig:instantiation-explicit-batcher}
\end{subfigure}
\caption{\tool's interface for explicitly-batched data structures and an instantiation}
\end{figure}

\tool provides the module signature shown in
\autoref{fig:explicitly-batched-interface} to describe
explicitly-batched structures.
This signature is a generalisation of the explicitly-batched interface
to \code{Counter} seen previously.  It uses the type \code{'a op} to
represent operations returning results of type \code{'a}, with
\code{wr_op} being used to tie each operation with a callback to
be called on completion. Requests to the data structure are then sent
through \code{run_batch}, which takes in batches of operations as
before and executes them.
A notable change to the original interface is the additional
\code{pool} parameter to \code{run_batch}, which is now provided so
that the implementation can also schedule tasks to be run on the
thread pool.

\begin{wrapfigure}[8]{r}{0.46\linewidth}
\vspace{-3pt}
\setlength{\abovecaptionskip}{5pt}
\setlength{\belowcaptionskip}{-5pt}
\begin{minipage}{1.0\linewidth}
\begin{minted}[fontsize=\footnotesize,
%
% linenos,numbersep=0pt, xleftmargin=7pt,
%
bgcolor=lgray]{ocaml}
module Make : functor (S : Batched) -> sig
  type t
  type 'a op = 'a S.op
  val init : pool -> t
  val apply : t -> 'a op -> 'a
end
\end{minted}
\end{minipage}
\caption{A functor for direct-style structure}
\label{fig:directstruct}
\end{wrapfigure}
Given a module \code{S} implementing the explicitly-batched
signature \code{Batched}, \tool provides a functor \code{Make} that can
construct a direct-style API to the data structure
(\autoref{fig:directstruct}). Clients of the data structure can then
use the \code{apply} operation to send operations to it, and the
direct-style interface means that client code can be written as if the
results are returned immediately.
Internally, when \code{apply} is called, \tool suspends the client
code via \code{await}, and uses channels to implicitly batch requests
to pass to the \code{run_batch} function, as hinted in
\autoref{sec:async} and will be detailed in \autoref{sec:dint}.

\begin{wrapfigure}[10]{r}{0.485\textwidth}
  \begin{minipage}{1.0\linewidth}
    \vspace{-1.5em}
\begin{minted}[fontsize=\footnotesize,
      % linenos,numbersep=7pt,xleftmargin=15pt,
      bgcolor=lgray]{ocaml}
module Counter = OBatcher.Make(BatchedCounter)
let pool = (* new thread pool *)
let c = Counter.init pool

let handle_request = fun _ -> 
  Counter.apply c Incr;
  printf "you are the %d'th visitor!" 
    (Counter.apply c Get)
\end{minted}
\end{minipage}
\caption{Using a direct-style batched counter}
\label{fig:countero}
\end{wrapfigure}
\autoref{fig:instantiation-explicit-batcher} presents the
straightforward instantiation of \tool's signature for our
explicitly-batched implementation of \code{Counter}.
Applying \tool's \code{Make} functor to this module then
returns a direct-style API, which we can integrate back into our
original request handler with minimal changes (\autoref{fig:countero}).
This new implementation, while providing the same API as the initial
version, is able to scale more gracefully with the number of
concurrent requests, making it a more apropos choice for a web-server.

In this way, \tool strikes a balance between a heavy-handed
coarse-grained locking around existing sequential code and fully
concurrent thread-safe re-implementations.
For more complex data structures, users can use \tool to gradually
improve the performance of code by incrementally exploiting more and
more high-level properties of the functions it exposes.
%

\begin{wrapfigure}[8]{r}{0.415\textwidth}
\vspace{-15pt}
\setlength{\belowcaptionskip}{-15pt}
\setlength{\abovecaptionskip}{5pt}
\begin{minipage}{1.0\linewidth}
  \begin{minted}[fontsize=\footnotesize,
    % linenos,numbersep=7pt,xleftmargin=15pt,
    bgcolor=lgray]{ocaml}
let handle_request = fun req ->
  if not (Set.apply seen_users
           (Insert (ip_addr req))) then
    Counter.apply c Incr;
    printf "you are the %d'th visitor!" 
           (Counter.apply c Get)
\end{minted}
\end{minipage}
\caption{Counting unique visitors}
\label{fig:uniq}
\end{wrapfigure}
For example, suppose we wanted to extend our endpoint to only record
\emph{unique} visitors using a set as shown in \autoref{fig:uniq}.
An explicitly-batched implementation of a concurrent set could
initially be written as a wrapper over a vanilla \code{Set}, using the
fact that membership queries are \emph{pure} to first handle all such
queries in a batch in parallel, and then sequentially process all
remaining requests.
From here, further improvements could be made by a more careful
analysis of the data structure, and implementing tailored
batch-parallel operations.



\section{Implementing \tool}
\label{sec:design}

In this section, we describe the technical details of \tool's
implementation in Multicore OCaml using asynchronous programming
primitives.
Then, observing that the design of the embedding is only dependent on
\code{async}/\code{await} functions, we demonstrate how the framework
can be ported to other languages by walking through and comparing to a
Rust implementation as an example.

\subsection{\tool in Multicore OCaml}
\label{sec:implicit}

The core of the \tool framework is the \code{Make} functor from
\autoref{fig:directstruct}, which is provided to the user to create their own
instances of concurrent data structures with a familiar direct-style
API, from explicitly-batched implementations that follow the
\code{Batched} signature---as we saw with the \code{Counter} example
in \autoref{sec:together}.
Internally, this functor, when instantiated, produces the following
three components for each explicitly batch-parallel data structure:
(1)~an extended type definition with a container for assembling
incoming asynchronous operations into batches, (2)~a direct-style
function \code{apply} that acts as the main interface for clients, and
(3)~a \code{try_launch} function that encodes the logic of how batches
are collected and forwarded to the underlying implementation via
asynchronous programming.
Below, we zoom in on the implementations of each one of the components (1)--(3).

\begin{wrapfigure}[6]{r}{0.36\linewidth}
\begin{minipage}{1.0\linewidth}
\vspace{-19pt}
\begin{minted}[fontsize=\footnotesize,bgcolor=lgray]{ocaml}
type t = {
  data: (* underlying data type *)
  container: Container.t;
  is_running: bool Atomic.t
  pool: Task.pool
  last_time: float }
\end{minted}
\end{minipage}
\label{fig:underlying-chan}
\end{wrapfigure}  
\subsubsection{Extended Data Type} 
The \code{Make} functor from \autoref{fig:directstruct} extends the
input data structure's representation \code{data} with additional
components: a \code{container} to collect the incoming requests for
operations, an atomic Boolean value \code{is_running} to indicate
whether a batch is being processed, a reference to the available
thread \code{pool} to submit the tasks for parallel execution, and the
time of the last batch execution.
This construction, where each instance of the data structure is given
a separate container to batch requests, is fundamentally what allows
\tool to have \emph{multiple} batched data structures in the same
system, adressing one of the main shortcomings of the original design
of implicit batching by \citet{AgrawalFSSU14}, discussed
in~\autoref{sec:intro}.
Any thread-safe channel with queue and dequeue functionality can be
used as a container for the batched operations, and \domainslib does
provide such a channel implementation. However, we note that we do not
need the elements inside to be ordered, nor do we ever need to dequeue
anything less than all elements. Hence, we implement a simple
variation of a thread-safe lock-free stack~\cite{Treiber:TR} to use as
container that strips away all but two operations: \code{push}, which
appends atomically to a list, and \code{pop_all}, which can be
implemented with a simple atomic \code{exchange} operation.

\begin{wrapfigure}[6]{r}{0.375\linewidth}
\vspace{-19pt}
\begin{minipage}{1.0\linewidth}
\begin{minted}[fontsize=\footnotesize,bgcolor=lgray]{ocaml}
let apply s op =
  let pr, set = Task.promise () in
  let req = Mk (op, set) in
  Container.send s.container req;
  try_launch s;
  Task.await s.pool pr
\end{minted}
\end{minipage}
\label{fig:apply}
\end{wrapfigure}
\subsubsection{Direct-Style Interface} 
\label{sec:dint}
The second component of the functor's construction is the function
\code{apply}, which is exposed as the main entry point for concurrent
client interactions with the data structure.
This function takes as its arguments an instance~\code{s} (of the
extended type~\code{t}) of the data structure and the operation's
metadata~\code{op}.
Internally, the \code{apply} function is implemented following the
intuition presented in~\autoref{sec:async} adapted for the conventions
of \domainslib.
The function first allocates an empty promise to capture the result of
the operation and acquires its resolution function; it then adds the
operation metadata and the resolution to the container.
Next, the function calls \code{try_launch}, potentially running the batch
itself, after which the function finally awaits on the initial promise.
We elaborate on \code{try_launch} below.

\begin{wrapfigure}[15]{r}{0.52\linewidth}
\vspace{-20pt}
\begin{minipage}{1.0\linewidth}
\begin{minted}[fontsize=\footnotesize,bgcolor=lgray]{ocaml}
let rec try_launch s =
  if has_no_requests t.container then () else
  let time = current_sys_time () in
  if not_enough_requests s.container &&
     time -. s.last_run < wait_threshold
  then Task.async s.pool (fun () -> try_launch s)
  else if Atomic.cas t.running false true 
  then begin
    let batch =
      Container.get_all s.container in
    s.last_run <- time;
    run_batch s.data s.pool batch;
    Atomic.set s.running false;
    Task.async s.pool (fun () -> try_launch s)
  end
\end{minted}
\end{minipage}
\setlength{\abovecaptionskip}{5pt}
\caption{The \code{try_launch} function}
\label{fig:try-worker}
\end{wrapfigure}  
\subsubsection{Launch Function} 
\label{sec:launch}
The final component of the functor is a helper function to
capture the logic of client tasks being temporarily promoted to
workers and handling batches of requests.
We described a simplified representation of this logic earlier in
\autoref{sec:async}, and we now present a more detailed, thread-safe
implementation that ensures a minimum batch size for better performance.
Client tasks can be promoted to workers whenever they observe that
there are enough pending requests and that no other worker is currently
running. To avoid a situation where a small number of pending requests
are never run, we also set a maximum duration that a batched data structure
can remain idle after the last batch run.
Should the minimum amount of pending requests not be met and
not enough time has passed since the last run, we do not simply exit
or spin, but instead we schedule another task to try launching the
batch again later. This ensures that we will not be deadlocked by a
small number of operations not being serviced and that the current
thread will not waste spin cycles waiting either.
Note that this design decision is optimised primarily for a
highly-concurrent worflkow where batches will accumulate quickly \ie a
web server; in a sequential setting, this particular encoding enforces
that each operation will take at least as long as the timeout, and a
user might instead consider an implementation such as one to be
presented in the next subsection without minimum batch sizes.
We also schedule another \code{try_launch} task at the end of each batch run,
to avoid another deadlock situation where other clients submit requests after
the existing worker has started running a batch.
%
  %
  


\subsection{Beyond OCaml: \tool in Rust}
\label{sec:rust-impl}
Though the narrative so far has only considered an OCaml
implementation of \tool, the reader may have noticed that the core of
this embedding does not require any language-specific features and
only depends on a select few asynchronous programming primitives.
Indeed, this captures one of the strengths of the framework: it is
\emph{lightweight} and can easily be generalised to other languages.
In the rest of this section, we substantiate this claim by presenting
a Rust implementation of the \tool framework and discuss any significant changes
that were needed.

\subsubsection{Explicitly Batched Interface}
Recall that the main client interface to \tool in the OCaml
implementation is its signature to describe explicitly-batched
data-structures.
While Rust does not support functors, we can recreate the essence of
this design using its trait mechanism.

\begin{wrapfigure}[14]{r}{0.375\linewidth}
\vspace{-15pt}
\begin{minipage}{1.0\linewidth}
\begin{minted}[fontsize=\footnotesize,bgcolor=lgray]{rust}
trait BOp { type Res; }

struct WrOp<Op : BOp>(Op,
   Box<dyn FnOnce(Op::Res) -> ()>);

pub trait Batched {
    type Op : BOp;

    fn init() -> Self;
    async fn run_batch(&mut self, 
        ops: Vec<WrOp<Self::Op>>)
      -> ();
}
\end{minted}
\end{minipage}
\setlength{\abovecaptionskip}{5pt}
\caption{Explicitly-batched interface}
\label{fig:rs-interface}
\end{wrapfigure}  
\autoref{fig:rs-interface} presents the embedding of \tool's
explicitly-batched interface into Rust.
The core of this encoding is captured in the \code{Batched} trait,
which is almost verbatim ported from the OCaml implementation.
In order to implement this trait for a given data-structure, the user
is required to supply three components:
(1) a type, \code{Op}, to describe the operations supported by the
data structure, (2) a function \code{init} to create empty instances
of the structure and finally (3) an \emph{asynchronous} function
\code{run_batch} which executes a batch of operations on the structure
and may exploit parallelism to optimise performance.
Note that as Rust's type system has no support for GADTs, we adopt a
more convoluted representation of operations and their return values
in this new encoding. In particular, operations are constrained
through a trait \code{BOp} with an associated type \code{Res} that
captures the return type of all operations. This is a less precise
encoding than before, as now the \code{Res} type must capture the all
possible return types of all operations, so clients must perform
additional case analysis to retrieve a usable result.
For example, to encode operations for the Counter from
\autoref{sec:overview}, we can define an enumeration type for
operations, \mintinline{rust}{enum CounterOp { Get, Incr }}, and
instantiate the \code{BOp} trait with optional integer as the result
type:
\mintinline{rust}{impl BOp for CounterOp { type Res = Option<i32> }}.

\subsubsection{Extended Data Type}
The next main component of the \tool implementation is the extended
data type definition that stores the additional metadata required to
implement batching.

The snippet on the right presents how this extended data-type
definition can be encoded in Rust.
The translation here from OCaml again is fairly mechanical.
We define a new struct \code{BatcherInner}, parameterised by an
explicitly batched datatype \code{B}. This struct maintains 4 pieces
of state: (1) an instance of the data structure, (2) a channel to
receive operations on, (3) a channel to send operations on, and (4) a
boolean to track whether any batch is currently running.

\begin{wrapfigure}[9]{r}{0.4\linewidth}
\vspace{-18pt}
\begin{minipage}{1.0\linewidth}
\begin{minted}[fontsize=\footnotesize,bgcolor=lgray]{rust}
struct BatcherInner<B : Batched> {
  data: Mutex<B>,
  recv: Mutex<Receiver<WrOp<B::Op>>>,
  send: Sender<WrOp<B::Op>>,
  is_running: AtomicBool
}

pub struct Batcher<B: Batched>(
   Arc<BatcherInner<B>>);
\end{minted}
\end{minipage}
\setlength{\abovecaptionskip}{5pt}
\end{wrapfigure}  
Finally, we wrap this inner data structure behind an atomic
reference-counted cell \code{Arc} to produce the final batched
structure \code{Batcher<B>}.
The most significant difference from the OCaml encoding arises from
Rust's more precise tracking of references and mutation, which
requires us to be more explicit about when references can copied (via
\code{Arc}) and what data is protected by locks (via \code{Mutex}).
Note that for simplicity we use separate locks for the structure and channel;
one could wrap both behind one lock to improve performance.


\begin{wrapfigure}[8]{r}{0.45\linewidth}
\vspace{-22pt}
\begin{minipage}{1.0\linewidth}
\begin{minted}[fontsize=\footnotesize,bgcolor=lgray]{rust}
async fn apply(&self, 
    op : B::Op) -> B::Op::Res {
  let (promise, set) = Promise::new();
  let wr_op = WrOp(op, Box::new(set));
  self.0.send.send(wr_op).unwrap();
  let s = self.clone();
  tokio::spawn(async move { 
    s.try_launch().await });
  promise.await }
\end{minted}
\end{minipage}
\setlength{\abovecaptionskip}{0pt}
\end{wrapfigure}
\subsubsection{Direct-Style Interface}
Having constructed this extended data type, we can now again recreate
a direct-style interface to our explicitly-batched data structures by
exploiting the \code{async}/\code{await} primitives provided by the
language to perform control inversion and reuse the same generic
pattern as used in our OCaml implementation. The code to do this is
presented in the snippet on the right.

This implementation is largely a verbatim translation of the OCaml
code (\cf~\autoref{sec:dint}).
%
%
In this case, the main changes that had to be made were to account for
slight discrepancies between the respective languages in their
treatment of asynchronous tasks and memory management.
In particular, asynchronous tasks in Rust are lazy, and will only be
evaluated when forced.
As such, in this implementation, when a client attempts to run a batch
of requests, it spawns a task (using \code{tokio::spawn}) which then
eagerly runs \code{try_launch} (to be discussed in the next section)
by \code{await}ing on it.
In conjunction to this, Rust's lack of a garbage collector now means
that the code must be explicit about when and where references are
shared, and in this case, an explicit \emph{clone} of the reference to
the batched data-structure is constructed and \emph{moved} into the
spawned task to use.

\begin{wrapfigure}[17]{r}{0.5\linewidth}
\vspace{-19pt}
\begin{minipage}{1.0\linewidth}
\begin{minted}[fontsize=\footnotesize,bgcolor=lgray]{rust}
fn try_launch(&self) -> BoxFuture<'static, ()> {
  let s = self.clone();
  async {
    if let Ok(_) = s.0.is_running
          .compare_exchange(false, true) {
      let mut recv = s.0.recv.lock().await;
      let mut ops = vec![];
      if recv.recv_many(&mut ops).await > 0 {
        let mut data = s.0.data.lock().await;
        data.run_batch(ops).await;
      }
      drop(recv);
      s.0.is_running.store(false);
      tokio::spawn(async move { 
        s.try_launch().await });
    }
  }.boxed() }
\end{minted}
\end{minipage}
\setlength{\abovecaptionskip}{5pt}
\caption{The \code{try_launch} function}
\label{fig:rs-try-worker}
\end{wrapfigure}  
\subsubsection{Launch Function} 
The final component required to complete this re-implementation of
\tool in Rust is an encoding of the \code{try_launch} operation which
will implement the logic of conditionally promoting clients to workers
and collecting and handling batches of requests.
\autoref{fig:rs-try-worker} presents the corresponding Rust
implementation of the \code{try_launch} function.
The core logic of this function is once again the same as in the
OCaml; for simplicity, this implementation omits any minimum
requirements on batch sizes and thus the associated timeout mechanism
they would require for liveness, but incorporating these would be
fairly trivial.
The main differences in the implementation arise due to subtleties of
Rust's treatment of asynchronous tasks.
In particular, Rust internally compiles asynchronous code into finite
state machines to optimise their execution.
As such, the compiler will reject an attempt to directly call
\code{try_launch} recursively as this induces infinite states in the
state machine, and so we must introduce a level of indirection and
explicitly allocate the future on the heap by boxing it.

Putting all together, as we have seen from this reimplementation in
Rust, the core of \tool's design is largely agnostic to the choice of
language, and the implementation can be fairly easily ported to any
language with suitable asynchronous programming primitives.
Generally speaking, we can expect that small changes may be needed to
account for slight discrepancies in the specific semantics of
asynchronous operations between languages, but otherwise, \tool
provides a general mechanism for introducing implicit batching into
new languages in a lightweight fashion.



\section{Engineering Batch-Parallel Implementations}
\label{sec:examples}

As demonstrated in \autoref{sec:overview}, \tool reduces the challenge
of implementing a concurrent data structure to implementing an
\emph{explicitly batch-parallel} one, which takes a collection of
operations and executes them in a way that maximises parallelism.
%
%
We argue that this design simplifies the task of engineering
thread-safe concurrency, relieving the developer from the need to
reason about concurrent interference, atomicity, and invariants, as
long as they can argue that the subsets of operations from the batch
provided to \code{run_batch} are safe to run in parallel---which only
requires understanding \emph{sequential} behaviour of those
operations.
The actual concurrent implementation is obtained by instantiating the
explicitly-batched interface \code{Batched} from
\autoref{fig:explicitly-batched-interface} with the type of the data
structure \code{t} and its operations \code{op}, constructor
\code{init}, and, most importantly, the function \code{run_batch} for
the parallel batch execution; the rest is handled by the \code{Make}
functor from \autoref{fig:directstruct}.

One can easily obtain a trivially correct batch-parallel
implementation from the sequential one by making \code{run_batch}
execute all given operations \emph{sequentially}, without any
parallelism whatsoever.
Such an \emph{ad-hoc} solution (\ie, the one that implements
\code{run_batch} explicitly) can be later refined into a ``more
parallel'' one, \eg, by separating all pure operations, to run them in
parallel, from the effectful ones, executed sequentially.
This highlights one of the main advantages of our methodology:
batch-parallel versions can be systemically derived from the original
sequential data structure.

One of this work's key observations is that, for certain classes of
sequential search structures, it is possible to identify implementation
strategies that further streamline the development of their
\emph{efficient} batch-parallel counterparts.
This section aims to showcase two such strategies, embodied by OCaml
functors: so-called \emph{split-join} (\autoref{sec:splitjoin}) and
\emph{expose-repair} (\autoref{sec:exposerepair}).
%
%
In \autoref{sec:adhoc} we present several ad-hoc batch-parallel
implementations that do not immediately fall under any of the above
two categories.
We note that the case studies we describe here are limited to search
data structures that implement maps and ordered sets (with the
exception of \autoref{sec:datalog}).
We do not claim that our implementations are optimal in the
theoretical sense, and leave such proofs for the future work.


\subsection{Split-Join Batching Strategy}
\label{sec:splitjoin}

The first batch parallelism strategy we explore is based on the idea
of \emph{splitting/joining} by \citet{BlellochFS16}. 
It is effective for various kinds of search structures represented as
\emph{balanced binary trees}, such as AVL trees, red-black trees, and
treaps, that allow one to divide them into multiple non-connected
trees that are themselves valid instances of that tree type. 
We can then perform insertion/deletion operations on each of these
sub-divided trees independently and in parallel, before joining them
together again to get the final tree after all operations have been
performed.
This idea has been previously explored in the works by
\citet{Akhremtsev2016} and by \citet{Sanders2019}, termed ``bulk
updates'' or ``bulk operations'', albeit not with the goal to derive
efficient concurrent data structures, but as an optimisation for
performing a large number of simultaneous updates on trees.
The novelty of this part of our work is, thus, in providing a
convenient abstraction to obtain efficient concurrent counterparts for
arbitrary search structures that fit this profile.

%

\begin{figure}[t]
\includegraphics[width=0.85\linewidth]{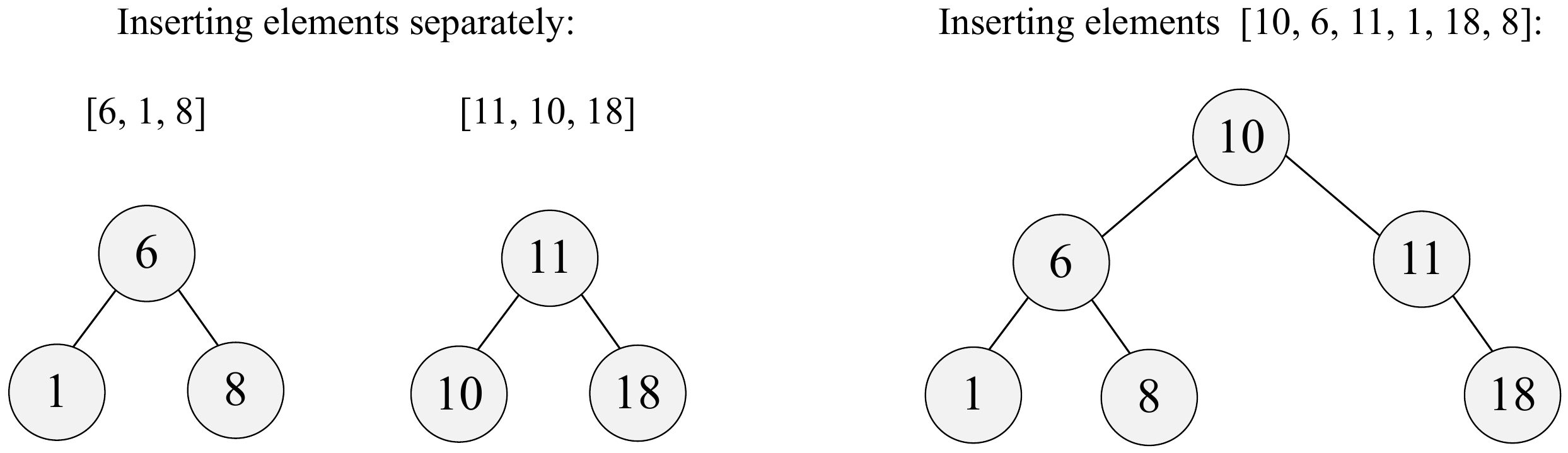}
\caption{Left: inserting elements into 2 empty separate trees. Right:
  inserting the same elements into a single empty tree (right).
  Joining the 2 trees on the left yields the single tree shown
  on the right, and splitting the tree on the right at element 11 
  yields the trees on the left.}
\label{fig:splitting}
\end{figure}


The crux of the approach is in splitting a tree instance in a way that
the combining the resulting sub-components after the parallel updates
can be done with better than $O(n)$ complexity.
In the case of binary search trees, a better complexity could be
achieved if we had to join trees pair-wise where the maximum key
element in the first tree, if any, is strictly less than the minimum
key element in the second tree, so the range of keys of these trees do
not overlap with each other. This would allow us not to explore those
trees in full when joining, as hinted by the example
in~\autoref{fig:splitting}, where most of the subtree structure
remains unchanged.
To achieve that, we adopt the idea from the work of
\citet{BlellochFS16}, who were themselves elaborating on the work of
\citet{Adams1993} on implementing elegant yet efficient functional
sets. They proved that \code{split} and \code{join} for balanced
binary trees, such as red-black trees, AVL trees, and treaps, can have
a worst-case time complexity of $O(\log n)$.
In this scheme, joining two balanced binary trees consists of
connecting them, comparing their balancing factor (\ie, \emph{height}
for AVL trees, \emph{black height} for red-black trees, and
\emph{priority} for treaps), and rebalancing them, if needed, by
disconnecting root nodes from child nodes and applying rotations.
Splitting applies a similar rebalancing strategy. 
Both these functions are called recursively only once per level, hence
their logarithmic complexity.

\subsubsection{A Functor for Split-Join Batching}
\label{sec:sjfunctor}

To provide a convenient abstract interface for split/join batch
parallelism, we define two modules signatures, \code{Sequential}
(\autoref{fig:split-join-sequential}) and \code{Prebatch}
(\autoref{fig:split-join-prebatch}), that outline the required
data types and functions to be provided by the user.

\code{Sequential} must contain the key type \code{kt} and the tree
type \code{'a t}, as well as an implementation of the constructor
\code{init} for creating a new data structure instance, \code{search},
\code{insert}, and \code{delete}. 
Note that operations that update the data structure should be done
in-place, hence the final return type of `unit`. We also do not
require \code{'a t} to be the type of the ``node''; it is often more
convenient to make it a \emph{wrapper over a root node} of a different
type that recursively contains child nodes of that node type. 
The users are expected to implement \code{Sequential} by defining all
basic sequential operations that can be used to interface with the
data structure. Should there be no need for batching, one can simply
invoke the functions here to use the data structure sequentially.

\begin{figure}[t]
    \begin{subfigure}{0.45\textwidth}
    \begin{minipage}{1.0\textwidth}
  \begin{minted}[fontsize=\footnotesize,linenos,numbersep=7pt,xleftmargin=15pt,bgcolor=lgray]{ocaml}
module type Sequential = sig
  type kt
  type 'a t    
  val init : unit -> 'a t    
  val search : kt -> 'a t -> 'a option
  val insert : kt -> 'a -> 'a t -> unit
  val delete : kt -> 'a t -> unit
end
\end{minted}
    \end{minipage}
    \caption{The \code{Sequential} module signature}
    \label{fig:split-join-sequential}
    \end{subfigure}
\begin{subfigure}{0.53\textwidth}
    \begin{minipage}{1.0\textwidth}
\begin{minted}[fontsize=\footnotesize,linenos,numbersep=7pt,xleftmargin=15pt,bgcolor=lgray]{ocaml}
module type Prebatch = sig
  module S : Sequential  
  val compare : S.kt -> S.kt -> int
  val set_root : 'a S.t -> 'a S.t -> unit
  val size_factor : 'a S.t -> int
  val split : 'a S.t -> S.kt -> 'a S.t * 'a S.t
  val join : 'a S.t -> 'a S.t -> 'a S.t
 end
\end{minted}
     \end{minipage}
    \caption{The \code{Prebatch} module signature}
    \label{fig:split-join-prebatch}
    \end{subfigure}
  \caption{\code{Sequential} and \code{Prebatch} module signatures the
    for split-join functor.}
\end{figure}
\code{Prebatch} is built on top of the \code{Sequential} module
signature, and describes additional functions that the split-join
functor will need to construct the batch-parallel version of the data
structure:
\begin{itemize}
\item The function \code{compare} exposes the comparison function for
  the key type. For example, if we were using \code{Int} as the key
  type, we can simply define it like so: \code{let compare =
    Int.compare}.
\item \code{set_root} swaps out the current root of the data
  structure. This is needed for updating the data structure in-place
  after joining its sub-components.
\item \code{size_factor} returns the number that over-approximates the
  size of the current data structure. For instance, this can be the
  height of a tree. We will describe its use shortly.
\item \code{split}: given a tree and a \emph{pivot} value~\code{k},
  returns two trees with non-overlapping key ranges where the maximum
  key of the first tree is strictly less than~\code{k}, and the
  minimum key of the second tree is equal or greater than~\code{k}.
\item \code{join}: provided two trees with non-overlapping key ranges
  and in ascending order based on those ranges, returns a single valid
  tree formed by joining the two input trees.
\end{itemize}

\begin{wrapfigure}[15]{r}{0.52\linewidth}
\vspace{-18pt}
\begin{minipage}{1.0\linewidth}
  \begin{minted}[fontsize=\footnotesize,linenos,numbersep=3.5pt,xleftmargin=13pt,bgcolor=lgray]{ocaml}
let run_batch t pool ops_array =
  let searches = ref [] in
  let inserts = ref [] in
  (* omitting deletions *)
  Array.iter (fun elt -> match elt with
  | Mk (Insert (key, vl), kont) ->
    (* safe to notify the client immediately *)
    kont (); inserts := (key,vl) :: !inserts
  | Mk (Search key, kont) ->
    searches := (key, kont) :: !searches
  ) ops_array;

  par_search pool t (Array.of_list !searches);
  par_insert pool t (Array.of_list !inserts)
\end{minted}
\end{minipage}
\caption{Split-join parallel batching}
\label{fig:sjbatch}
\end{wrapfigure}  
%
%
For each data structure that allows for efficient implementation of
the split and join operations, the user need only implement these two
modules and their functions. 
Once done, \tool's split-join functor can take over and define an
explicitly-batched module, much like the one we
%
%
have seen in
\autoref{sec:together}.
Let us examine its \code{run_batch} function defined in
\autoref{fig:sjbatch}. 
For any given batch of operations, we start by separating different
types of operations. We currently limit our implementation to search
and insert operations, but it should be a fairly mechanical process to
extend the functor to accommodate other effectful operations as well,
following the handling of either batched searches or inserts as a
template. 
After separation, we execute all searches, then all insertions. Even
if we had interleaving search and insert requests in our initial
batch, executing the operations this way is still correct from the
perspective of linearisability, as we have discussed previously with
the \code{Counter} example in \autoref{sec:ebatcher}. 
There is no particular pre-processing needed for batch parallel
searches as those do not modify the data structure, and we simply use
\domainslib's higher-order \code{parallel_for} function to dispatch
the search operations in parallel.


\begin{figure}[t]
  \begin{subfigure}{0.55\textwidth}
  \begin{minipage}{1.0\textwidth}
\begin{minted}[fontsize=\footnotesize,linenos,numbersep=5pt,xleftmargin=15pt,bgcolor=lgray]{ocaml}
let par_insert ~pool s inserts =
  let n =
    Array.length inserts / seq_threshold + 1 in
  (* assume that inserts are randomly ordered *)
  let pivots =
    Array.init n (fun i -> fst inserts.(i)) in
  sort pivots;
  let s_arr = split_multiple s pivots in
  let sub_ranges =
    partition pivots inserts in
\end{minted}
  \end{minipage}
  \end{subfigure}
  \begin{subfigure}{0.43\textwidth}
  \begin{minipage}{1.0\textwidth}
\begin{minted}[fontsize=\footnotesize,linenos,numbersep=5pt,xleftmargin=15pt,bgcolor=lgray,firstnumber=11]{ocaml}
parallel_for pool
  ~start:0
  ~finish:(Array.length sub_ranges)
  ~body:(fun i ->
    let range = sub_ranges.(i)
    for j = fst range to snd range do
      let k, v = inserts.(j) in
      S.insert s_arr.(i) k v
    done);
set_root t (join_multiple s_arr)
\end{minted}
\end{minipage}
\end{subfigure}
\caption{Batch-parallel insertion in the split-join functor.}
\label{fig:par_insert}
\end{figure}

Parallel execution of inserts is a bit more subtle; it relies on the
various \code{Prebatch} functions defined earlier. 
We begin by checking whether (a)~the present size of the data
structure warrants splitting by invoking \code{size_factor} and
whether (b)~the number of insert operations is sufficiently high. We
set some threshold for these two, with the intuition that smaller data
structure sizes and smaller numbers of operations are not worth the
extra overhead of splitting, parallelising, and rejoining, in which
case we will simply perform the operations sequentially (we omit this
part from our presentation).
Otherwise, we perform the parallel insert procedure as shown
\autoref{fig:par_insert}. We select a number of random pivots based on
the size of the batch and the sequential threshold (lines~2-7), and
use them to split the tree (line~8) and partition the insert batch
into ordered sub-batches (lines~9-10). We then use \domainslib's
\code{parallel_for} to perform each sub-batch of operations on its
respective subtree in parallel (lines~16-19). Finally, we rejoin the
split trees together (line~20).
Our implementation takes advantage of some other opportunities for
parallelisation, namely, sorting the random pivots via a classic
implementation of parallel merge sort, and partitioning the insertion
operations with a parallel partition procedure inspired by the one
used in the QuickSort algorithm.

%
%

\subsubsection{Case Study: Red-Black Tree}

As a concrete example of the split-join strategy for batch parallelism
and its respective functor in action, let us take a look at the
red-black (RB) tree~\cite{Bayer72}, a classic self-balancing binary
search tree data structure.
In addition to searching and insertion of elements, its typical
implementations support efficient deletion, minimum, and maximum---all
of them enjoying logarithmic worst-case time complexity.


\begin{wrapfigure}[9]{r}{0.40\linewidth}
\vspace{-15pt}
\includegraphics[width=0.95\linewidth]{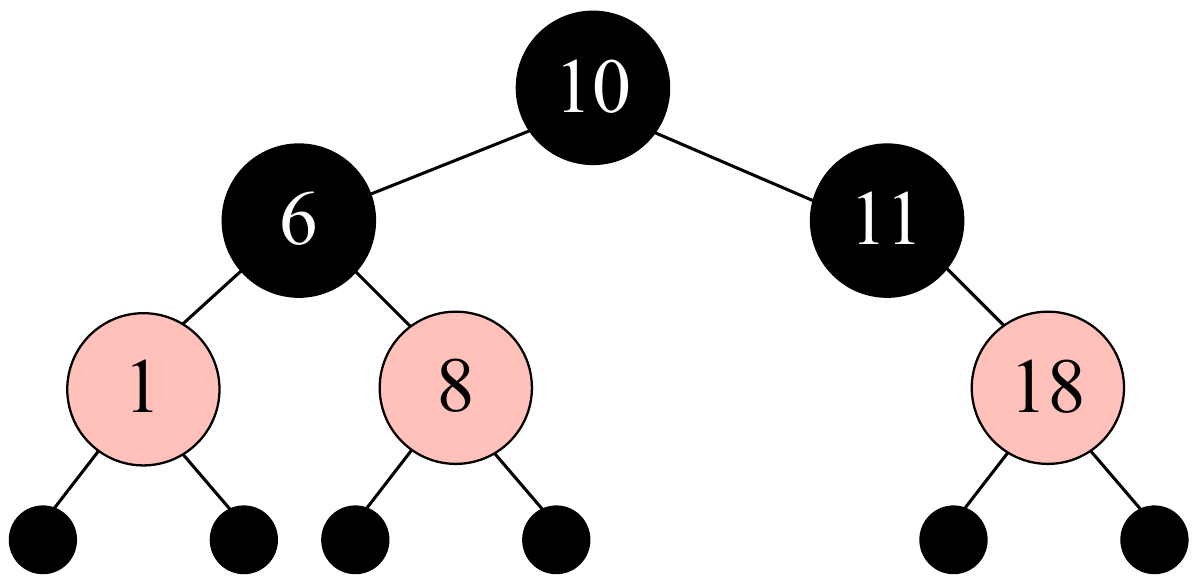}
\caption{Example of a red-black tree}
\label{fig:rbtree}
\end{wrapfigure}  
\paragraph{Sequential red-black tree overview} 
A red-black tree is an \emph{approximately} balanced binary search
tree, meaning that it is not perfectly balanced, but instead
guarantees that no root-to-leaf path \emph{is more than twice} as long
as any other root-to-leaf path. 
This is achieved by assigning each node a colour, which can be either
red or black, and ensuring that the following invariants are upheld:
(1)~every leaf (not containing any key) is black, (2)~if a node is
red, both its immediate children must be black, and (3)~each path from
a given node to any leaf must have the same black height, \ie, the same
number of black nodes.
Some presentations also add an extra condition: the root node must be
black. We will omit this rule for our implementation, as it is not
essential for the desired time complexity, and we must allow a red
root for the split and join functions later on. 
Aside from this omission, our implementation follows the red-black
tree algorithms as described in the standard textbook by \citet{clrs}.
\autoref{fig:rbtree} shows the balanced binary tree that we've
previously seen in \autoref{fig:splitting}, but as a red-black tree,
with some of its nodes are coloured in red, so that the tree 
invariants stated above are respected. 

Following the \code{Sequential} signature from
\autoref{fig:split-join-sequential}, whose effectful functions (\eg,
\code{insert}) modify the data structure in-place, we implement a
wrapper type \code{'a t} around the root node of the red-black tree.
Searching a red-black tree follows the same procedure as for an
unbalanced binary search tree. Inserting a node involves recursively
traversing the tree like the search operation and adding a new node at
the leaf level, after which, the tree might be repaired by recolouring
and rotating as needed (we refer the reader to Chapter~13 of the
textbook by \citet{clrs} for the details).




\paragraph{Batch-parallel red-black tree overview} 

The logic of a batch-parallel red-black tree follows the general
split-join strategy from \autoref{sec:sjfunctor}, requiring one to
implement the \code{set_root}, \code{size_factor}, \code{split}, and
\code{join} functions to obtain a fully functional batch-parallel
version of the data structure.

The \code{set_root} function is straightforward: we simply replace the
root node in the red-black tree wrapper type. 
For the sake of other functions, we
augment our sequential implementation with an additional piece of
information stored in each node: the \emph{black height}, \ie, the
number of black nodes on the paths from that node to each
leaf~\cite{BlellochFS16}.
%
%
This will serve as the basis for the \code{size_factor} function: we
can use the black height as an estimate of the size of the tree.
This addition also enables implementations of \code{split} and
\code{join} operations.
For these functions, we faithfully recreate the algorithms as
described in pseudocode by \citeauthor{BlellochFS16}, where
rebalancing and recolouring occur depending on the black height
difference between the input trees. Storing black heights avoids the
additional $O(\log n)$ cost of counting the number of black nodes
every time we need this information, hence keeping \code{split} and
\code{join} to $O(\log n)$ time complexity as well.

\subsubsection{Other Split-Join Data Structures}

As shown, balanced binary search trees are especially well-suited for
the split-join batch-parallel paradigm, subject to their own
\code{split} and \code{join} functions.
The work by \citet{BlellochFS16}
also provides blueprints of these functions for other types of search
structures: AVL trees~\cite{avl} and treaps~\cite{SeidelA96}.

Like an RB tree, an AVL tree balances itself through rotations.
It does not have any colour code, and instead preserves the invariant
that at any given node in the tree, its left and right sub-trees have
a height difference of at most one.
%
%
A \emph{treap}, meanwhile, is a probabilistically balanced tree where
each node is randomly assigned a priority number, and we rotate the
tree after each update to ensure that the priority number of each node
is higher than the prioriy numbers of its child nodes, essentially,
recreating a max heap based on the random priority number.


We implemented their batch-parallel versions just like with the RB
tree by defining their \code{split} and \code{join} functions as
described by \citet{BlellochFS16}.
To implement the \code{size_factor} function, we simply use the tree
height already stored in AVL tree nodes, and augment the treap nodes
with stored tree heights as well.
Our accompanying code provides the respective working implementations.


\subsection{Expose-Repair Batching Strategy}
\label{sec:exposerepair}

The split-join strategy discussed in \autoref{sec:splitjoin} is
beneficial for the search structures that admit sublinear-time changes
in their shape without breaking their invariants, such as rebalancing.
For instance, we rely on being able to move nodes around in binary
search trees to split and rejoin them.
Not every search structure lends itself well to being
batch-parallelised this way, meaning that they cannot be split into
sub-structures of the same shape, or that joining them would induce a
prohibitively large overhead undoing the performace improvement gained
from parallelism.

This is the case for search structures, such as bitwise tries, that
use string/binary prefixes or hashes to store key values: in those
structures the \emph{position} of an element in the data structure
\emph{is} its key.
Therefore, ``physically'' splitting such a structure tree will
inevitably require one to deal with complex re-indexing logic.
%
%
On the flipside, we can argue that such a search structure is even
better suited to a batch-parallel implementation since each update
should only affect its predictable, \emph{localised} area that is the
same regardless of what other keys populate the data structure.

%
%

Making such ``position-based'' search structures batch-parallel might
not be as straightforward as it seems. 
Advanced examples like the \citeauthor{VanEmdeBoas1977}
tree~(\citeyear{VanEmdeBoas1977}), the x-fast trie, and the y-fast
trie~\cite{Willard1983}, which we will showcase for this section,
contain additional metadata, pointers, or even entire secondary
sub-structures to achieve the promised sub-logarithmic time complexity
of their operations in a sequential case.
Those sub-structures must be accounted for, restored, and/or updated
both before and after running a batch.

In this section, we present a novel strategy for batch parallelism,
dubbed \emph{expose-repair}, which is aimed to address such search
structures. 
It centres around first ``exposing'', or preparing the structure
before each batch of operations, so as to make sure parallel
operations in each its localised area do not affect each other. Then,
we ``repair'' the result after processing the batch of operations, to
re-establish the metadata or sub-structures covering the entire data
structure.

\subsubsection{Expose-Repair Functor Overview}
\label{sec:exposerepairoverview}

\begin{figure}[t]
\begin{subfigure}{0.45\textwidth}
\begin{minipage}{1.0\textwidth}
\begin{minted}[fontsize=\footnotesize,linenos,numbersep=4pt,xleftmargin=12pt,bgcolor=lgray]{ocaml}
module type Sequential = sig
  type kt
  type t
  val init : int -> t
  val mem : t -> kt -> bool
  val insert : t -> kt -> unit
  val delete : t -> kt -> unit
  val predecessor : t -> kt -> kt option
  val successor : t -> kt -> kt option
end
\end{minted}
    \end{minipage}
    \caption{The \code{Sequential} module signature}
    \label{fig:expose-repair-sequential}
    \end{subfigure}
    \begin{subfigure}{0.53\textwidth}
    \begin{minipage}{1.0\textwidth}
\begin{minted}[fontsize=\footnotesize,linenos,numbersep=4pt,xleftmargin=12pt,bgcolor=lgray]{ocaml}
module type Prebatch = sig
  module S : Sequential
  type dt
  val compare : S.kt -> S.kt -> int
  val expose :
    S.t -> S.kt array -> S.kt array * dt
  val repair : S.t -> dt -> unit
  val insert_range :
    S.t -> S.kt array -> dt -> int * int -> unit
 end
\end{minted}
\end{minipage}
\caption{The \code{Prebatch} module signature (simplified)}
\label{fig:expose-repair-prebatch}
\end{subfigure}
\caption{\code{Sequential} and \code{Prebatch} module signatures for
  the expose-repair functor.}
\end{figure}

Similarly to the split-join functor from \autoref{sec:splitjoin}, for
expose-repair we define two modules signatures, \code{Sequential} and
\code{Prebatch} that need to be instantiated by the user. For
simplicity, we phrase the \code{Sequential} interface as a \emph{set}
rather than as a key/value, \ie, it only stores the keys of the type
\code{kt}, following the presentation from the standard
textbook~\cite[Chapter~20]{clrs}.
The key/value map-like functionality can be easily restored by
associating the ``satellite data'' with the keys.
As the result of this design, the \code{Sequential} signature shown in
\autoref{fig:expose-repair-sequential} offers the \code{mem} function
instead of \code{search}.
The signature also features the \code{predecessor} and
\code{successor} functions, as one of the main selling points of the
fast search structures, such as van Emde Boas tree, the x-fast and the
y-fast tries, which all offer the $O(\log\log u)$ time complexity for
these operations (where $u$ is the largest integer key that can be
stored in the structure).
Just like in the case of split-join strategy, the pure query
operations like \code{mem}, \code{predecessor}, and \code{successor}
can be dispatched in parallel in a separate stage of the batch
processing, without any special preparation.

The signature for the \code{Prebatch} module is best explained in
parallel with the implementation of a parallel executor for the
effectful operations in the respective expose-repair functor provided
by \tool, which makes use of the \code{Prebatch} components.
As a characteristic example, consider the implementation of parallel
insertion via expose-repair strategy shown
in~\autoref{fig:par_insert_expose_repair}.
It starts by checking if the number of insert operations in the batch
exceeds the sequential threshold, and if it is below, performs the
operations sequentially (we omit this part from the listing for
brevity); otherwise, its continues with the the batch processing.
To do so, it first obtains \code{n}~sorted random ``tentative pivots''
from the batch of operations, and transforms them into pivots we can
use to partition our batch of operations (lines 2-7).
Unlike the split-join functor, which takes random elements from the
batch of operations as pivots on which to split the trees
(\autoref{sec:sjfunctor}), in the case of position-based structures we
would want pivots that can separate the data structure into logical
parts consistent with their inner layout and the recursive workings of
their respective sequential operations.
For instance, in the case of the van Emde Boas tree, those pivots
would be \emph{multiples of the square root of the size of its
  universe} (\ie, the set of all its possible keys), so that we can
divide up a cluster of trees without dissecting any tree in the middle
(more details are given in~\autoref{sec:vebtree}).
Next, the implementation prepares the data structure for batch
processing where needed and partitions the batch of operations using
the obtained pivots (lines~7-8).
The preparation can take the form of (but is not limited to)
pre-initialising parts of the data structure, or temporarily removing
some pointers between nodes belonging of different sections of the
data structure. This process might both rely on and/or inform the
creation of the pivots.
The partitioning (line~8) only depends on the values of the pivots and
the arguments of the insert operations, hence is not
structure-specific.
Next, the sub-batches of operations resulting from the partitioning
are run in parallel on the same pre-processed data structure
(lines~9-10).
It is done in the assumption that \code{expose} has indeed returned
the pivots in a way that split the range of the insertions in a way so
that operations in different sub-batches would not interfere with each
other.
Finally, the initial data structure \code{t} is repaired by, \eg,
removing unused pre-initialised parts, updating metadata, and
restoring auxiliary pointers between its sub-parts (line~11).

\begin{figure}[t]
\begin{minipage}{1.0\linewidth}
  \begin{minted}[fontsize=\footnotesize,linenos,numbersep=5pt,xleftmargin=15pt,bgcolor=lgray]{ocaml}
let par_insert ~pool t inserts =
  (* omitted: checking if the batch size is larger than seq_threshold *)
  let n = Array.length inserts / seq_threshold + 1 in
  (* assume that inserts are randomly ordered *)
  let pivot_seeds = Array.init n (fun i -> fst inserts.(i)) in
  sort pivot_seeds;
  let pivots, dt = expose t pivot_seeds in
  let sub_batch_ranges = partition pivots inserts in
  parallel_for pool ~start:0 ~finish:(Array.length sub_batch_ranges)
    ~body:(fun i -> insert_range t inserts dt sub_batch_ranges.(i));
  repair t dt
\end{minted}
\end{minipage}
\caption{Batch-parallel insertion in the expose-repair functor.}
\label{fig:par_insert_expose_repair}
\end{figure}

The code in \autoref{fig:par_insert_expose_repair} relies on the
following components of the \code{Prebatch} functor from
\autoref{fig:expose-repair-prebatch}:

\begin{itemize}
\item An abstract type \code{dt}, which is specific for each
  particular search structure and is used to store supplementary
  information for batching, if needed (\cf~\autoref{sec:vebtree} for a
  concrete example).
\item The function \code{expose} takes as its input the data structure
  itself and an array of ``tentative pivots''. Having those, it
  (a)~prepares the data structure in-place for batch processing and
  (b)~returns an array of the pivot values used to partition our batch
  of operations.
  We combine these two procedures, as they can be very much
  intertwined with one another.
\item The function \code{repair} repairs the data structure after all
  updates have finished.
\item Finally, \code{insert_range} performs a sequence of insertions
  into the exposed data structure sequentially. 
  It takes as input the data structure, a reference to the whole batch
  of insertions with the range of the sub-batch (to avoid reallocating
  a new array for each sub-batch), and extra information in the form
  of~\code{dt}. Note that \code{insert_range} takes the whole data
  structure as input, since inserting a specific range of values
  should keep the effects localised and therefore should not affect
  any other instance of \code{insert_range} running in parallel.
\end{itemize}
Conceptually, the correctness of an expose-repair strategy is simpler
than split-join for trees.
Broadly speaking, one can argue for the correctness of an
implementation from the following assumptions about the operations: 1)
first, \code{expose} should preserve the elements in the structure and
should only logically split the structure into disjoint ``chunks'' via
pivots, 2) second, for a given range, the \code{insert_range} function
should not affect ``chunks'' outside of it, and finally 3)
\code{repair} should not lose elements and restores any data structure
invariants.
%
%
%
We illustrate this in more detail in our argument for the correctness
of the van Emde Boas Tree in the next section.
Supporting parallel deletions using the expose-repair strategy would
require implementation of a function with a signature similar to
\code{insert_range}, which we have omitted for the sake of brevity.

\subsubsection{Case Study: Van Emde Boas Tree}
\label{sec:vebtree}


The van Emde Boas tree (vEB tree for short) represents a priority
queue and was conceived as a way to resolve bottlenecks in in query
operations (\eg, membership, predecessor, successor) for ordered,
random access sets~\cite{VanEmdeBoas1977}. 
When implemented as binary trees, such as RB tree, these operations
have $O(\log n)$ time complexity.
The vEB tree instead allows for membership, predecessor, and successor
queries to be done in $O(\log\log u)$ time, where $u$ is the universe
size, or the number of all possible key values the vEB tree might
store. This logarithmic speedup over balanced binary trees is achieved
at the cost of space efficiency, with the vEB tree occupying $O(u)$
space regardless of how many elements its contains.

\begin{wrapfigure}[8]{r}{0.36\linewidth}
\vspace{-16pt}
\centering
\includegraphics[width=0.78\linewidth]{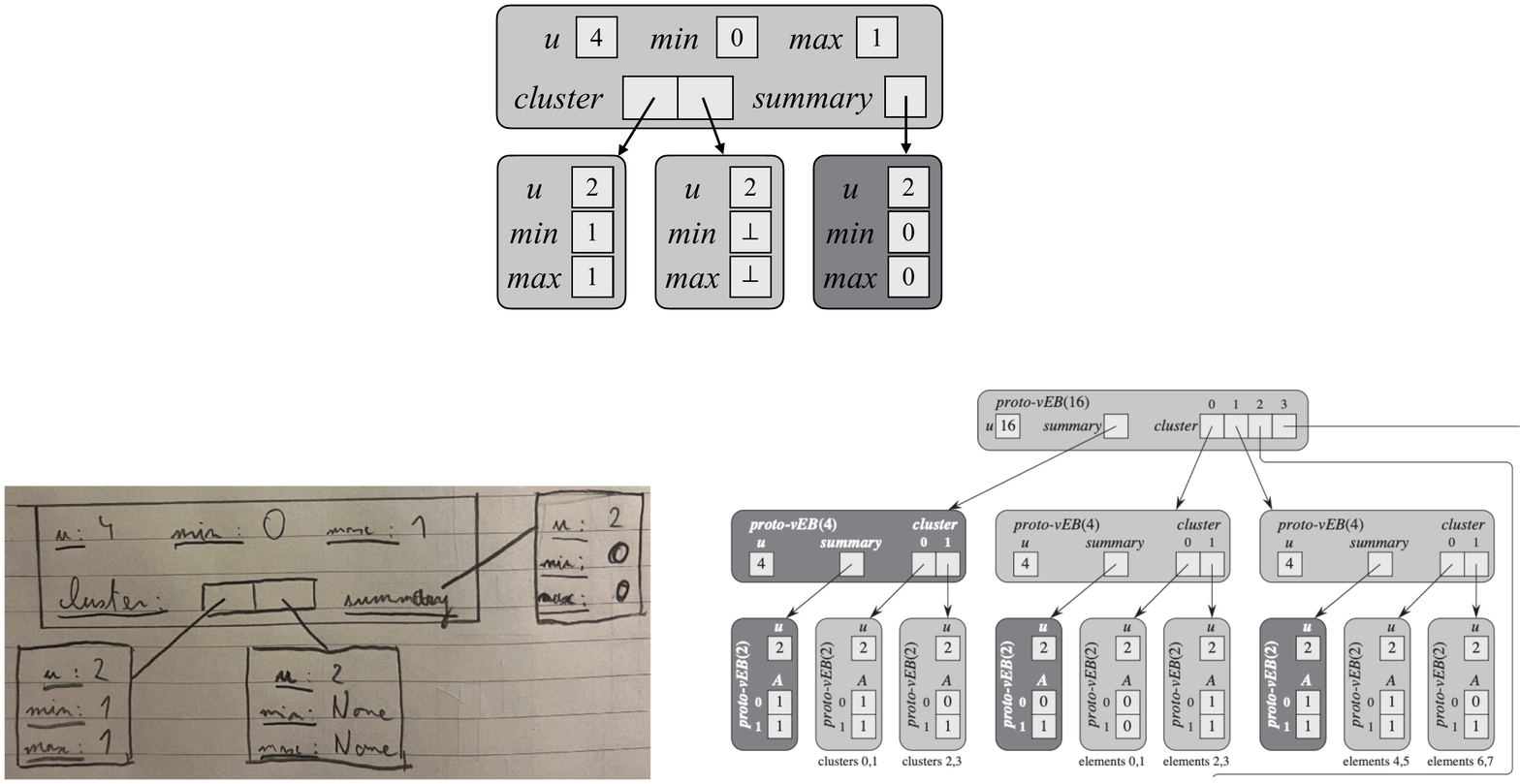}
\setlength{\belowcaptionskip}{-10pt}
\setlength{\abovecaptionskip}{5pt}
\caption{A vEB tree containing $\set{0, 1}$.}
\label{fig:vebtree}
\end{wrapfigure}  
\paragraph{Sequential van Emde Boas tree overview} 
A vEB tree is structured recursively, with each node containing
(1)~the universe size $u$ at that node, (2)~the minimum value of the
tree at that node, (3)~the maximum value of the tree at that node,
(4)~an array \code{cluster} of $\sqrt{u}$ vEB tree nodes, each of
which contains a tree of universe size $\sqrt{u}$ so that a vEB tree
of an index~$i$ would contain keys in the range
$[i\sqrt{u}, (i + 1)\sqrt{u} - 1]$, and (5)~a summary vEB tree of size
$\sqrt{u}$ storing the indices of vEB trees in the cluster array that
are non-empty, \ie, have at least one element.

Consider the example vEB tree in~\autoref{fig:vebtree}.
It has the universe size of~4, so its possible key values range from~0
to~3. 
In this case, it only contains the keys~0 and~1: 0 is stored in the
\emph{minimum} field of the root node and is thus is not present in
the vEB nodes below (the minimum field of a vEB node is not just
metadata, but always contains the key itself).
The \emph{maximum} field of the root vEB node shows 1, but this is
\emph{not} the key value itself, but is just metadata.
Looking down, the left child node, corresponding to the index 0 of
the root cluster, itself has no cluster, and both its minimum and the
maximum fields store~1.
It is a \emph{base case} vEB node, whose universe size is 2, and whose
maximum field \emph{can} contain the key itself.
%
%
The vEB node at the root cluster index~1 is empty, as shown with both
its minimum and maximum being~$\bot$.
Finally, the root features an additional \emph{summary} vEB node,
which just contains 0, meaning that only the vEB child node at the
position 0 of the root cluster is non-empty.
Though it may seem odd in this example to ``summarise'' a cluster of
size 2, for larger clusters it allows one to find the first non-empty
cluster in $O(\log \log u)$ time, enabling fast predecessor and
successor queries.
We omit detailed descriptions of sequential vEB operations, and refer
the reader to Chapter~20 of~\citet{clrs} or to our
accompanying implementation.

\paragraph{Batch-parallel van Emde Boas tree overview} 

The vEB tree lends itself naturally to batch processing. 
We note that since there are $\sqrt{u}$ trees available in the cluster
at the root level, for any reasonable choice of universe size, there
will be a sufficient number of vEB trees even at the first level, that
we need not look further below for parallelising our batched
operations on the vEB tree.

\begin{wrapfigure}[11]{r}{0.4\linewidth}
\vspace{-15pt}
\begin{minipage}{1.0\linewidth}
  \begin{minted}[fontsize=\footnotesize,numbersep=5pt,bgcolor=lgray]{ocaml}
let expose t arr =
  let size_cluster =
    lower_sqrt t.uni_size in
  let pivots =
    Array.init
      (Array.length arr)
      (fun i ->
        high t arr.(i) * size_cluster)
  (dedup pivots, ())
\end{minted}
\end{minipage}
\setlength{\abovecaptionskip}{5pt}
\caption{The vEB tree \code{expose} function}
\label{fig:veb-expose}
\end{wrapfigure}
The code of the \code{expose} function for vEB tree is shown in
\autoref{fig:veb-expose}.
We will not need any supplementary information, so the type \code{dt} is
just \code{unit}. 
We do however need to identify how to create appropriate pivots from
random keys in our batch of insert/delete operations in order to split
the vEB tree into logical chunks. For this, we simply transform each
pivot key into the minimum possible key for its target sub-vEB tree in
the root-level cluster array, which is done using a standard helper
function \code{high}
%
%
taken from \citet{clrs} (Chapter 20).
This ensures that keys in different sub-batches are never inserted in
the same vEB tree, nor in the same cluster sub-array. At the end, we
deduplicate the final list of
pivots, as two different keys going to the same vEB tree will yield
the same pivot.
For each key \code{k} to be inserted in the sub-batch of operations
defined by the input range, \code{insert_range} invokes the sequential
insertion procedure, but \emph{does not} update the summary node, the
minimum, or the maximum at the root level.
This last update is exactly what is done by the respective
\code{repair} function.

With respect to the correctness of this implementation, we find that
it follows the conditions stated at the end of
\autoref{sec:exposerepairoverview} and thus yields a correct vEB
at the end of batched operations:

\begin{enumerate}
  \item The implementation of \code{expose} for the vEB tree does not
  modify the data structure at all, but merely returns the ranges for
  non-overlapping cluster sub-arrays.
  \item Since we made pivots such that keys in different sub-batches
  cannot be inserted in the same vEB tree or even the same cluster
  sub-array, each \code{insert_range} call cannot affect anything
  other than its own ``chunk'' of the vEB tree.
  \item The \code{repair} function restores metadata at the root
  node by updating the minimum, maximum, and summary tree. This is
  the only we need to update at the end, as each cluster node vEB
  tree is already valid at the end of parallel insertion.
\end{enumerate}

\subsubsection{Other Expose-Repair Data Structures}

\begin{figure}[t]
\begin{subfigure}{0.48\textwidth}
\begin{minipage}{1.0\textwidth}
  \begin{center}
    \includegraphics[width=0.85\linewidth]{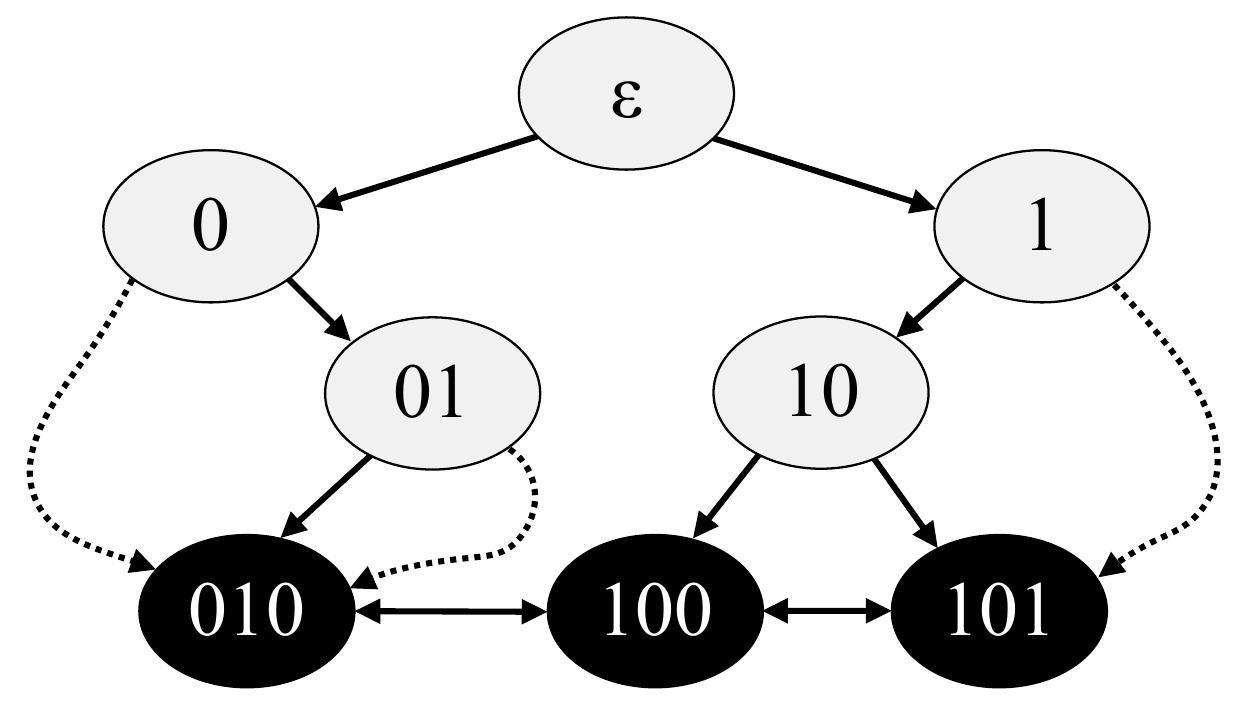}
  \end{center}
\end{minipage}
\caption{The x-fast trie}
\label{fig:xfast-trie}
\end{subfigure}
\quad
\begin{subfigure}{0.47\textwidth}
\begin{minipage}{1.0\textwidth}
  \begin{center}
    \includegraphics[width=0.75\linewidth]{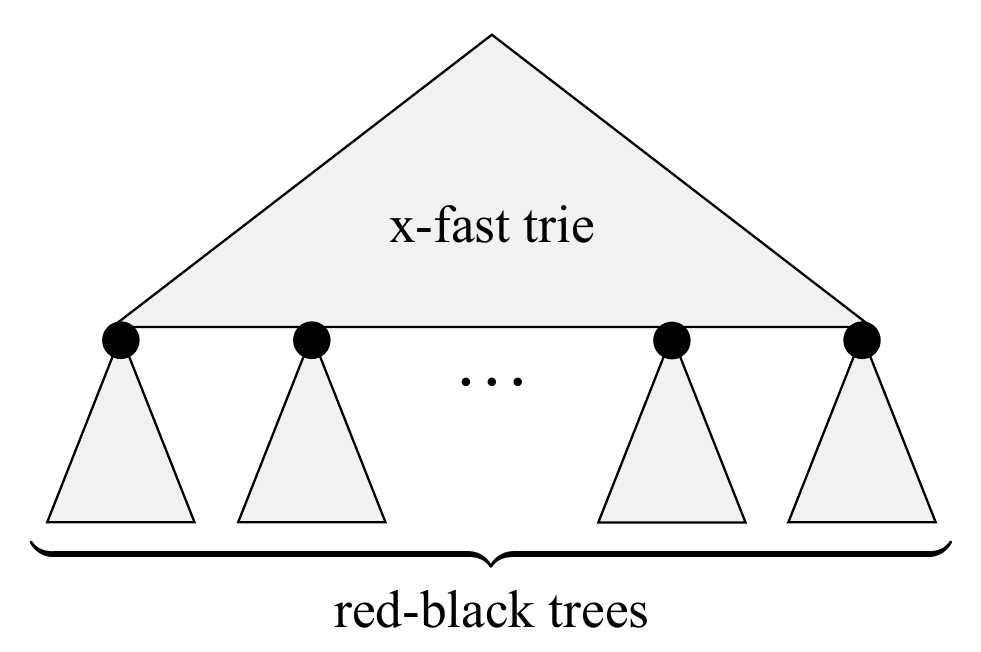}
  \end{center}
\end{minipage}
\caption{The y-fast trie}
\label{fig:yfast-trie}
\end{subfigure}
\setlength{\abovecaptionskip}{5pt}
\setlength{\belowcaptionskip}{-10pt}
\caption{X-fast trie and Y-fast tries. For the x-fast trie in
  \autoref{fig:xfast-trie}, solid double arrows indicate pointers from
  each leaf to the previous and the next leaf, while dashed single
  arrows indicate descendant pointers.}
\end{figure}

In addition to vEB, we have instantiated the expose-repair for two
more position-based search structures: the x-fast and y-fast trie by
\citet{Willard1983}.
Both these structures were designed to preserve the query
time complexity of the vEB while only using $O(n\log u)$ and $O(n)$
space respectively, where $n$ is the number of elements in the trie.



In an x-fast trie (\autoref{fig:xfast-trie}), all full unsigned
integer keys are stored at the same leaf level, with there being as
many intermediate layers as there are bits representing these
integers.
%
%
Each leaf node points to its successor and/or predecessor leaf node.
Each internal node with no left child contains a pointer to the
smallest leaf in its right subtree, similarly, each internals node
with no right child contains a pointer pointer to the largest leaf in
its left subtre; in both cases those are referred to as
\emph{descendant} pointers.
Typically, at each layer an x-trie has a hash table to accelerate
queries. 
%
%
As a proof of concept, our implementation uses arrays, which worsens
the worst-case x-fast trie's space complexity,
but allows for a relatively simple expose-repair implementation: we
\code{expose} the sub-trie by removing pointers and adding
intermediate nodes up until the layer determined by the number of
pivots, and partition our operations among the nodes at that layer.


To turn the x-fast trie into a y-fast trie (\autoref{fig:yfast-trie}), we replace the simple leaf nodes with a forest of red-black trees.
%
The size of each RB tree has an expected size of $\log u$, and it
is split as needed to maintain that size.
%
%
Implementing a batch-parallel version of the y-fast trie using the
expose-repair functor is not very complicated.
%
%
For insertions, \code{expose} determines the ranges for the keys to be
inserted and RB trees that each parallel task should cover based on
random pivots from the batch of operations, dispatching those
operations sequentially within \code{insert_range}.
At the end, \code{repair} checks the size of each resulting RB tree,
splitting them where needed to respect the size bound.




\subsection{Case Studies with Ad-Hoc Batch Parallelism}
\label{sec:adhoc}

We conclude this section by presenting two implementations of ad-hoc
batch parallelism that do not immediately follow either of the
split-join or expose-repair strategies: a B-tree and a stateful
Datalog solver.
We have additionally implemented a batch-parallel version of a skip
list~\cite{pugh1990skip} by adopting the design proposed
by~\citet{AgrawalFSSU14}; a curious reader can find its high-level
description in \autoref{sec:skiplist}.


\subsubsection{B-Tree}
\label{sec:btree}
\begin{wrapfigure}[9]{r}{0.47\linewidth}
\vspace{-13pt}
\setlength{\belowcaptionskip}{-15pt}
\setlength{\abovecaptionskip}{8pt}
\includegraphics[width=1.0\linewidth]{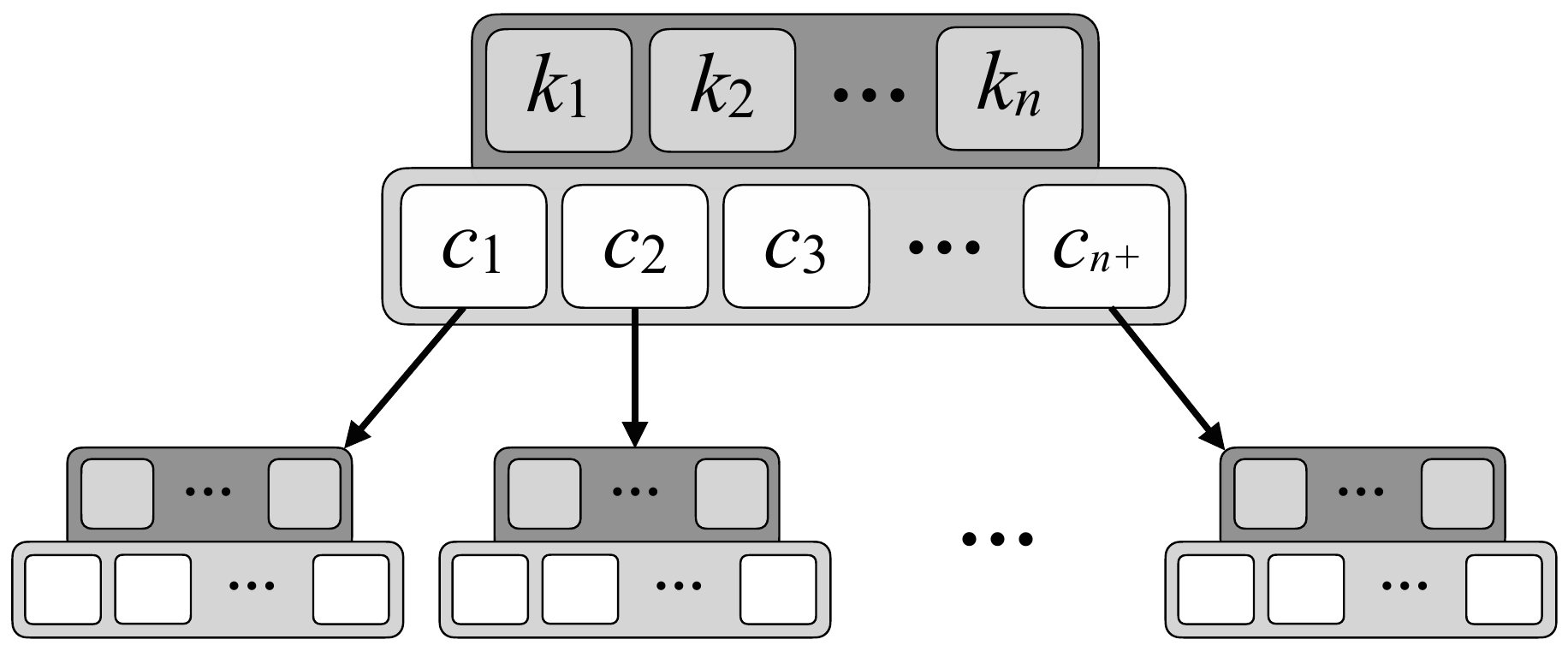}
\caption{Structure of a B-tree}
\label{fig:btree}
\end{wrapfigure}

A B-tree is a widely-used tree data structure that can be used to
implement an efficient set or map interface~\cite{bayer1972symmetric}.
The B-tree's efficiency arises from the fact that its operations are
carefully structured to always maintain the balance of the tree and
try to preserve a good distribution of values across the tree, thereby
ensuring a logarithmic lookup time.
\autoref{fig:btree} depicts the general structure of a B-tree.
The nodes in a B-tree contain 2 components: a sorted sequence of keys
${k_1, \dots, k_n}$ stored in the node, and a sequence of sub-trees
$c_1, \dots c_{n+1}$ corresponding to the node's children; leaf nodes
contain keys, but no children.
The B-tree forms a search tree 
%
%
in the fact that the keys in a node
define \emph{intervals} that bound the contents of children: a child
$c_i$ will only contain keys that lie in the range
${k_{i-1} \leq v < k_i}$.
Finally, to provide a consistent performance,
the B-tree enforces the constraint that for any node, all of its
sub-trees must all have the same height (\ie, the tree is balanced),
and that every interior node will have a number of keys between
$t - 1$ and $2t - 1$, where $t$, the branching factor, is a global
parameter of the tree, ensuring that values are evenly distributed.

%

Adding new values to a B-tree requires care to preserve the its
invariants.
The high-level intuition for an insert is to recursively traverse down
the tree to find the leaf into which to insert the value---but what if
the leaf is already at full capacity (\ie, it contains $2t - 1$ keys)?
To solve this, insertion relies on the \emph{splitting} operation:
when a particular node in the data structure has reached full
capacity, we can gain additional capacity while preserving the
structure's invariants by splitting the node into two and adding a new
key and child to the node's parent.
While splitting requires modifying the parent, we can prevent this
from bubbling up the entire B-tree by maintaining an invariant that
the node being visited during the recursive traversal
has \emph{at least one free space}, thus, preemptively splitting a
sub-tree before descending down the tree.
In this way, even if inserting a value into a node requires splitting,
we can be guaranteed that \emph{no further splits} higher up in the
tree would be required to restore the B-tree's intrinsic invariants.

%

%
\paragraph{A batch-parallel B-tree}
Implementing a batch-parallel search in B-tree is straightforward:
when processing a batch of searches, first dispatch the values that
occur at the current node, partition the remaining values by the
children, and recurse on each sub-tree in parallel.
%
%
%
At a high level, the same divide-and-conquer strategy should be
applicable to B-tree inserts: when inserting a batch of values into a
B-tree, we can partition the values by sub-trees and recurse on each
child in parallel.
The problem with this strategy is in ensuring that the nodes of the
B-tree never become over-full during the inserts, as this causes them
to split, possibly causing a cascade of splits to ``bubble up'',
making them interfere with other parallel operations.

%
%
%
Inserting $n$ elements into a given node can require splitting up to
$n$ children in the worst case, \ie, when each of the elements falls
into a \emph{different} child, each of which itself needs splitting.
If we only look at the contents of a single node, then the maximum
capacity of any node will be at most $t$: this follows from the fact
that the B-tree requires that every node has a number of keys between
$t-1$ and $2t - 1$, so starting from the minimum and encountering $t$
splits would already cause the node hit the maximum number of keys.
As the size of an input batch is typically far larger than $t$, a
constant parameter of the tree, this overly-conservative bound would
hamper the parallelism.

\begin{wrapfigure}[10]{r}{0.475\linewidth}
  \begin{minipage}{1.0\linewidth}
\setlength{\abovecaptionskip}{0pt}
\vspace{-15pt}
{\footnotesize    \begin{align*}
      \text{nkeys}(n) = & ~ \text{number of keys in the node $n$} \\
      \text{child}(n,i) = &~ \text{$i^\text{th}$ child of the node $n$} \\
      \text{free}(n)  = &~ 2 t - 1 - \text{nkeys}(n) \\
      \text{cap}(n)  = &~ \begin{cases}
            \text{free}(n)  & \text{$n$ is a leaf} \\
                                \begin{matrix}
            \left(1 + \min_i \text{cap}(\text{child}(n, i))  \right) \times \\
               \text{free}(n)
                                \end{matrix}
                            & \text{otherwise} \\
      \end{cases}
    \end{align*}
}  \end{minipage}
\caption{Capacity of a B-tree}
\label{fig:new-capacity}
\end{wrapfigure}
The key insight in our design is in defining a novel notion of
capacity for a B-tree node (\cf~\autoref{fig:new-capacity}) that is
more precise but still relatively cheap to compute.
%
%
%
%
For a leaf node, the capacity is simply the number of elements that
can be inserted before it hits max capacity. For an interior node,
notice that each time one of its children splits, it increases the
number of keys that the node has by 1, until it reaches max-capacity
at which point an additional element could cause the node to require
splitting.
As such, we can conservatively bound the capacity of a node by the
capacity of its children plus 1, that is, the minimum number of
elements to cause a child to split, multiplied by the free slots it
has, the number of times it can allow its children to split.
By using the minimum here, we lose some precision in our bound, but
this also means that the value of this statistic can be computed
without imposing significant overhead.

Using this new definition of capacity, we can construct a
batch-parallel B-tree insert using the divide-and-conquer strategy.
In particular, after partitioning the batches by the children, if the
size of a batch exceeds the capacity for the sub-tree, then we
preemptively split the sub-tree until the desired capacity is reached.
Once all the splits have been performed, we can then recursively
insert into the children in parallel, returning to the standard divide
and conquer strategy.
%

\begin{wrapfigure}[7]{r}{0.35\textwidth}
\vspace{-1.75em}
\begin{minipage}{1.0\linewidth}
\begin{minted}[fontsize=\footnotesize,bgcolor=lgray]{ocaml}
let run_batch t pool ops =
  let searches, inserts = 
     (* partition batch *) in
  sort pool searches; 
  sort pool inserts;
  par_search t pool searches;
  par_insert t pool inserts
\end{minted}
\end{minipage}
\end{wrapfigure}
Putting it all together, we incorporate both these search and insert
implementations to instantiate the explicitly batched interface of
\tool and construct an efficient batch-parallel B-tree (right).
We first partition the input batch, splitting out the searches and
inserts to be handled separately.
These partitions are then pre-sorted using an off-the-shelf parallel
merge-sort implementation.
Finally, we execute the requested batch by first dispatching the
searches using our batch-parallel search implementation, and then
subsequently dispatching the inserts using the corresponding
batch-parallel insert one.


\subsubsection{Datalog Solver}
\label{sec:datalog}

\begin{wrapfigure}[6]{r}{0.40\textwidth}
\vspace{-18pt}
\begin{minipage}{1.0\linewidth}
\begin{minted}[fontsize=\footnotesize,bgcolor=lgray]{ocaml}
module Datalog : sig
  type db
  type term 
  val insert: db -> term -> unit
  val query: db -> term -> term list
end
\end{minted}
\end{minipage}
\end{wrapfigure}
As our final case study, we demonstrate the generality of \tool by
investigating its use for constructing a \emph{thread-safe} and
\emph{efficient} wrapper over a popular OCaml Datalog
library~\cite{ocaml-datalog} with minimal additional implementation
effort.
%
%
The main interface to the library is through a type \code{db} encoding
Datalog databases is presented on the right.
Here, each database \code{db} represents a collection of facts known
to be true, where each fact is a relation applied to zero or more
symbols, \eg, \code{edge("a","b")} might encode knowledge that an edge
exists between nodes \code{"a"} and \code{"b"} in some graph.
%
%
Users can extend a database using the \code{insert} operation,
submitting new facts or adding deductive rules that describe how to
deduce new facts from existing ones.
%
%
%
%
%
%
Once a database has been constructed, \code{datalog} allows users to
{query} the set of facts in the database.
By submitting terms \code{query}, \ie, \code{edge("a","b")} or
\code{connected("x",y)}, the database will return a list of all facts
that match the query, answering questions such as \emph{is there an
  \code{edge} between \code{"a"} and \code{"b"}?} or \emph{what nodes
  are connected to~\code{"x"}?}
%
%
%
The \code{datalog} library
%
%
was not written with a multicore environment in mind.
It makes pervasive use of mutation and shared state that would be
challenging to port to a fine-grained implementation.
In contrast, implementing a batched interface to the library is
comparatively simple: when handling a batch of requests, we first
dispatch all queries in parallel, and then evaluate each of the
inserts sequentially.
We use this simple idea to instantiate \tool's interface with the
\code{datalog} library.







\section{Evaluation}
\label{sec:evaluation}

The main intended use for concurrent search structures is to allow
concurrent tasks to exchange data without introducing communication
bottlenecks.
Since such structures form the majority of our case studies, in
\autoref{sec:evalsearch} we provide an extensive evaluation of their
throughput trends.
%
%
In \autoref{sec:evaldatalog}, we evaluate the performance of the
concurrent Datalog solver.
The goal of our experiments is not to claim the maximum performance,
which still requires carefully crafted concurrency; instead, we aim to
show that our approach provides reasonable performance with less work.

%

All the reported benchmark results were obtained by running the
experiments on an AWS EC2 \texttt{c7i.12xlarge} server instance
equipped x86 Intel$^\text{\textregistered}$
Xeon$^\text{\textregistered}$ Scalable (Sapphire Rapids) processor
with 24~physical cores and 96~GB of memory, running Ubuntu 22.02 with
OCaml 5.1.1.


%

\subsection{Benchmarking the Concurrent Search Structures}
\label{sec:evalsearch}



\begin{figure}[!h]
\centering
\vspace{5pt}
\input{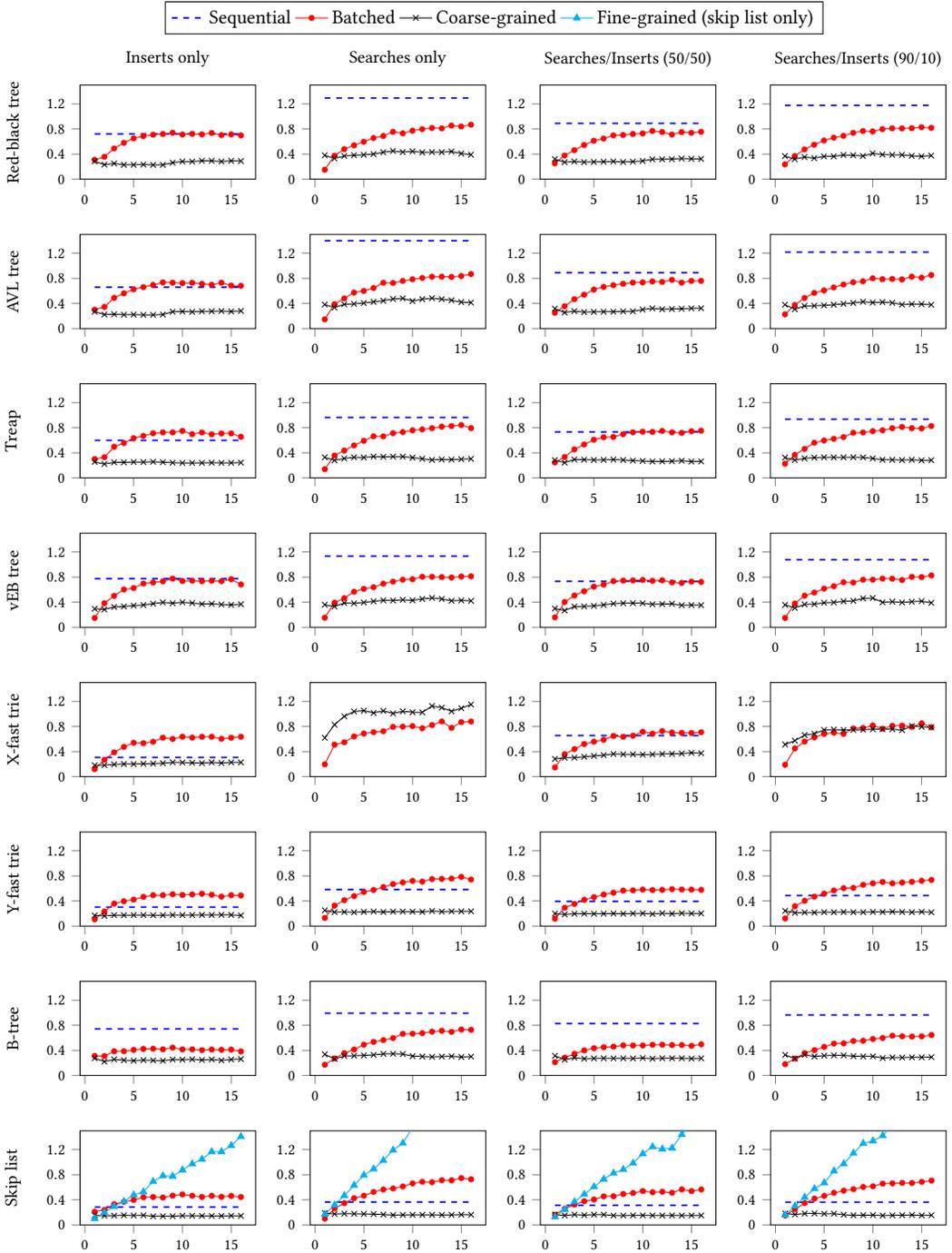}
\caption{Throughput comparison for concurrent search structures from
  \autoref{sec:examples}.
  The x-axes represent the numbers of involved domains/cores, and
  y-axes show throughput in millions of operations per second.}
\label{fig:functor-perf-results}
\end{figure}


We evaluate the throughput for all eight search structures from
\autoref{sec:examples} (RB tree, AVL tree, treap, van Emde Boas tree,
x-fast and y-fast tries, B-tree, and skip list) on the same set of
benchmarks.
For each benchmark, we fix the number of initial elements in the data
structure at 2,000,000 and the workload size to be
1,000,000 operations. 
We experiment with four different workload setups: inserts only,
searches only, 50\%/50\% and 90\%/10\% search/insert split.
Each operation of the workload is submitted to \domainslib's thread
pool as a separate concurrent task.
Each data point takes the average of five runs, performed after five
warm-up runs. 
We compare the performance of our batch parallel implementations with
their respective coarse-grained and sequential implementations,
varying the number of domains -- thin wrappers over system-level
threads provided by the OCaml runtime.
\autoref{fig:functor-perf-results} shows our results, allowing us to
draw several observations detailed below.

\paragraph{General performance trends}
%
First, we observe that in nearly every benchmark, our batch-parallel
generated search structures outperform their coarse-grained
counterparts by a significant margin starting from two domains, with
the gap widening as we increase the number of domains.
The gap is particularly evident for insertion-heavy benchmarks, where
we see the batched implementations match or even outperform the
sequential ones when more domains are available.
Fully sequential executions typically outperform the batch-parallel
ones for searches, even for higher domains and by a significant
margin. This is likely due to the efficiency of searches (even when
done sequentially), combined with the comparatively large overheads
for creating/managing parallel tasks.

Second, we observe that in nearly all cases, in the case of a single
domain, a batch-parallel implementation shows a lower throughput than
the coarse-grained one. 
This is because with a single domain, the batch-processing routine is
launched for every submitted operation after the waiting period
between batches in the \code{try_launch} function
(\autoref{fig:try-worker}), due to not meeting the minimum batch
size---as there are no other threads contributing operations.
%
Therefore, we conclude that, without further optimisation,
batch-parallelism only pays off in strictly multi-threaded scenarios.

Third, we observe that our x-fast and y-fast tries are in general
less performant than our binary trees (AVL and RB), despite their
superior asymptotic time complexity, even when considering only the sequential
implementations.
We conjecture that this is due to their much heavier memory use than
that of the binary trees, since each key requires \emph{multiple
  nodes} to represent, and so adding nodes would take more time
putting additional stress on OCaml's concurrent memory allocator.

Finally, we note that almost all the batch-parallel data structures
show reasonable speed-ups: the throughput increases with the number of
domains, up through approximately 8 domains. From that point onwards,
throughput grows more slowly with the number of domains. This slowing
growth at larger numbers of domains presumably stems from the
increased synchronisation overhead.





\paragraph{Concurrent searching in an x-fast trie}
%
The x-fast trie search-only and 90/10 search-insert split benchmarks
are the only scenarios in our suite, where our batched implementation
struggles to even match the coarse-grained implementation. This is
the only data structure where search time is $O(1)$.
%
Indeed, the sequential benchmark throughput for searching in an x-fast
trie is orders of magnitude higher than of the other data structures,
exhibiting around $22 \cdot 10^6$ op/sec in the search-only benchmark,
and around $2.8 \cdot 10^6$ op/sec for the 90/10 search-insert split
benchmark.
We can surmise that this is a rare case where the underlying operation
is so fast that, for a small number of threads, the lock contention
imposes a smaller overall overhead than starting a batch.
%

\paragraph{Comparison with a fine-grained skip list}
%
We have also compared our batched skip list implementation to the
fine-grained lazy skip list by \citet{HerlihyLLS07}, which we have
ported from Java to OCaml.
Sadly, \autoref{fig:functor-perf-results} clearly indicates the
growing performance gap between the fine-grained and the
batch-parallel implementation.
However, instead of simply admitting the crushing defeat by the giants
of the multiprocessor programming, we are going to make a few
observations that should encourage the reader to not dismiss batching
as a methodology for implementing concurrent structures.
While both implementations measure comparable amount of code---200 LOC
for the fine-grained skip list and 300 LOC for the batch-parallel
one---the fine-grained implementation is significantly more intricate
in its design.
Specifically, \citeauthor{HerlihyLLS07}'s skip list makes use of
\code{Atomic} references and introduces locks associated with each
node so that the nodes can be physically removed in a ``lazy''
fashion---the pattern known to be difficult to reason about formally
when proving linearisability~\cite{Vafeiadis08}.
In contrast, the batch-parallel implementation does not feature any
synchronisation primitives whatsoever, relying exclusively on
\code{parallel_for}.
It should also be noted that the batch-parallel implementation can
also benefit from various sequential optimizations. In the context of
skip lists, preprocessing steps, such as sorting inserts or removing
k-smallest elements are analytically more efficient than their
iterative counterparts~\cite{HendlerIST10}. More generally,
deduplication of requests could also stand to improve throughput. Our
current implementation only uses pre-sorting as an optimisation but
could be improved further.
%
%



\subsection{Benchmarking the Concurrent Datalog Solver}
\label{sec:evaldatalog}

\begin{wrapfigure}[10]{r}{0.42\textwidth}
\vspace{-12pt}
\centering
\begin{tikzpicture}[scale=0.60]
\pgfplotsset{
xticklabel style={
        /pgfplots/xtick pos=bottom,
        /pgfplots/xtick align=outside,
},
yticklabel style={
        /pgfplots/ytick pos=left,
        /pgfplots/ytick align=outside,
},
}
    \begin{axis}[
      height=5cm,width=8cm,
      legend style={at={(0.45,1.4)},anchor=north,font=\small, legend columns = -1},
      ylabel = {Throughput (ops/sec)},
      title = {(90/10) Queries/Inserts},
      scaled y ticks={base 10:-4},
      ymin=0, ymax=200000,   
      ytick distance=50000
      ]
\legend{Batched, Coarse-grained, Batched (Sequential)};
\addplot[mark=otimes*, color=red, mark size=2.2pt]
table [x=domains, y=datalog-batched-throughput, y
error=datalog-batched-sd, col sep=comma]              
{data/aws16/datalog_results.csv};
\addplot[mark=x,color=black, mark size=2.7pt]                         
table [x=domains, y=datalog-coarse-throughput, y
error=datalog-coarse-sd, col sep=comma]                
{data/aws16/datalog_results.csv};
\addplot[mark=square*,color=cyan, mark size=2.0pt]      
table [x=domains, y=datalog-non-parallel-batched-throughput, y
error=datalog-batched-sd, col sep=comma] 
{data/aws16/datalog_results.csv};

\end{axis}
\end{tikzpicture}
\setlength{\abovecaptionskip}{5pt}
\caption{Datalog solver throughput.}
\label{fig:datalog-perf-results}
\end{wrapfigure}
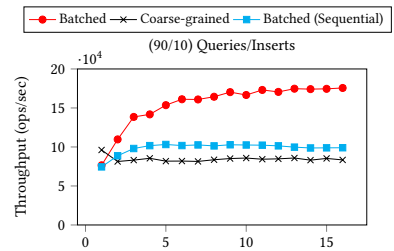
For our experiment with the \code{datalog} library, we implement a
typical graph-connectedness constraint system on top of the Datalog
database, using a graph of 200 nodes, with 30,000 initial edges, and
submit requests that either add new edges or query about the
connectedness of two nodes in the graph.
All implementations are evaluated on a workload of 50,000 concurrent
requests, using a standard 90\%/10\% split between queries and
inserts, varying the number of domains/cores and taking the average of
10 runs for each configuration.
\autoref{fig:datalog-perf-results} presents the results of
this evaluation.
As can be seen, our construction is efficient and scales gracefully as
the number of domains increases, while the corresponding
coarse-grained wrapper's performance gradually degrades as more
domains are added, suffering from increasing lock contention.
Our plot also includes the throughput of a na\"{i}ve batch-parallel
implementation of a \code{datalog} wrapper, which simply executes all
operations, queries and inserts alike, in a sequence.
Interestingly, its throughput still outperforms the coarse-grained
version, showing the benefits of the amortised cost of locking
provided by the batched interface yet it is unable to exploit
parallelism.



\section{Related Work}
\label{sec:related}

\paragraph{Flat combining and delegation-based locking}
The technique of flat combining itself draws from long-running line of
research into delegation-based locking.
These works all investigate minimising lock contention by delegating
concurrent requests to be executed under a dedicated worker.

\citet{oyama1999executing} were the first to introduce this strategy,
and considered using a simple LIFO queue to keep track of additional
requests that may have accumulated while a client was in a critical
section.
\citet{HendlerIST10} developed the flat combiner by extending this
design and pre-allocating empty entries in the task queue for each
thread thereby minimising contention between clients when appending
tasks.
To avoid situations where the queue is mostly empty, the flat combiner
design also incorporates strategies for periodically compacting the
queue and only keeping entries from threads that have recently
submitted tasks.
Subsequent works have built on this design and investigated optimising
the configurations for factors such as cache
locality~\cite{FatourouK11,RoghanchiEB17}, and specialised multicore
hardware~\cite{LoziD0LM16}.
Most recently, \citet{GuptaDKPZK23} developed a technique to
transparently incorporate delegation-based locking into existing
systems by constructing a dynamic lock that manipulates the call stack
at runtime to suspend the client and send arbitrary critical section
code to be executed by a dedicated worker, allowing delegation to be
introduced into a system without significant source-level changes.

Delegation-based locking research has primarily focused on mitigating
lock contention and, as such, prior work has not considered exploiting
parallelism within the worker itself as in \tool.
The insight in this work is that these techniques from
delegation-based locking can be combined with
\code{async}/\code{await} primitives to support implicit-batching in a
lightweight fashion.

\paragraph{Batch parallelism and data parallelism}
Data parallelism has long been a common approach for implementing
parallel computations, and often it is implemented in a ``batched''
manner. For example, if you have a function that needs to be applied
to a large number of data records, it is quite natural to divide the
records into batches to apply the function in parallel.

Batch-parallel data structures do indeed bear a resemblance to
data-parallel collections available as libraries, \eg, in
Haskell~\cite{JonesLKC08} and Scala~\cite{ProkopecBRO11}.
The main difference between data parallelism and batch parallelism is, 
perhaps, subtle.
The former approach focuses on optimising the execution of \emph{one}
aggregate-style bulk operation (\eg, \code{map}, \code{reduce},
\code{filter}, \etc.) at a time by exploiting possibilities for
parallelism allowed by the structure itself.
From the perspective of the user a call to a data-parallel \code{map}
is the same as a call to a sequential \code{map}, which just works
faster.
Importantly, neither of those calls assume concurrent interference
from other calls made to the same collection. Finally, for data
parallelism, it is typically easy for the programmer to construct
explicit batches, as the parallelism to exploit has a simple structure.
%


By contrast, batch parallelism solves the problem of running
\emph{multiple} concurrent operations to a data structure in the most
efficient way, while being aware of possible interference between
them. Batch parallelism operates on structured data, extracting
parallelism from whatever operations happen to arrive, on the fly.
Moreover, the benefit of implicit batch parallelism is that it can
handle situations where the batch structure is not, a priori, clear or
easy to determine.
That said, for each particular such operation in its batch, a
batch-parallel implementation can make use of data-parallel
techniques, even though we did not exercise this possibility in the
scope of this work.

\paragraph{Other batch-parallel data structures} 

Over the last several years, researchers have developed a variety of
batch-parallel data structures that focus on the problem of how to
process a batch of data structure updates in parallel
without being concerned how the batches are being constructed in the
first place.
%
%
The existence of these types of batch-parallel data structures
provides good motivation for the creation of a generic implicit
batching library like the one provided by \tool.
%

As of today, batch-parallel structures and algorithms are an active
research area.
\citet{TsengDB19} developed dynamic trees that could support batch
update operations, and then used them to maintain Euler tours.
\citet{AcarABDW20} also explored the problem of maintaining dynamic
trees subject to batches of updates, handling a broader class of
queries. \citet{TsengDS22} have developed algorithms for maintaining a
minimum spanning forest, which they used for graph clustering.
\citet{WangYGS21} developed a batch-parallel data structure for
solving the closest pair problem.

There has also been significant work on batch-parallel graph
algorithms in the Massively Parallel Computation (MPC) model. For
example, \citet{ItalianoLMP19} show how to process batch updates to
graphs, while supporting connectivity queries and while maintaining a
minimum spanning tree or a matching. \citet{GilbertL20} generalise
this for the $k$-server model, a different model of parallel
computation. \citet{DhulipalaDKPSS20} provide yet more general results
for maintaining dynamic graphs subject to batch updates in the MPC
model.


\paragraph{\tname{Batcher}}

The design of \tool is inspired by \citeauthor{AgrawalFSSU14}'s
\tname{Batcher}~(\citeyear{AgrawalFSSU14})---a scheduler implemented
as a part of modified Cilk-5 runtime thats allows to make batching
implicit in a way that is transparent to the user.
Rather than being a lightweight solution based on provided programming
language mechanisms, \tname{Batcher} fundamentally is an extension of
a standard randomised work-stealing scheduler that allows operations
to a data structure to implicitly form a batch before being sent for
processing in an asynchronous setting.
For context, it is important to note that the main contribution of
\citeauthor{AgrawalFSSU14}'s work is primarily on a theoretical side:
their aim is to provide a scheduler with a provable complexity bound
on an execution time of a program that manipulates a batch-parallel
data structure, in terms of its \emph{work} (the total number of
operations the program must execute) and \emph{span} (the length of
the longest non-parallelisable sequence of operations).
The statement of such a performance theorem, if possible, would be
further complicated in the case of multiple concurrently used
batch-parallel data structures, hence \tname{Batcher} only supports
one such structure.
%
%
In contrast, \tool supports multiple data structures by design, but
comes with no formal guarantees regarding its performance.


\paragraph{Concurrent search structures}  

In this work, we have implemented a variety of concurrent search
structures. For some of these data structures, such as B-trees,
Red-Black trees and AVL trees, there is a long history of concurrent
implementations~\cite{Ellis,Bayer77}, as well as more recent work,
particularly on B-trees and related
structures~\cite{Braginsky12,Srivastava22}.
Most relevant to this paper is the work by
\citet{Blelloch22,BlellochFS16,Blelloch98}, who showed how splitting
and joining trees yielded very efficient parallel solutions---our
implementation of the split-join paradigm followed this approach.
 
Search structures with sub-logarithmic time, such as the van Emde Boas
tree, x-fast tries, and y-fast tries, have been much more challenging
to implement in concurrent settings, and there exist relatively
few such parallel/concurrent implementations.
Of note, the only concurrent implementation (to our knowledge) of an
x-fast (or y-fast) trie is the SkipTrie by~\citet{Oshman13}. The first
parallel version of the van Emde Boas tree (to the best of our
knowledge) was just published by~\citet{Gu23}. More recently,
\citet{Khalaji0DA24} developed a very impressive concurrent van Emde
Boas tree implementation relying on Intel's Hardware Transactional
Memory instructions.
A related sub-logarithmic data structure was developed
by~\citet{Brown20}, providing a very efficient concurrent
implementation of an Interpolation Tree, not considered in this paper.

\section{Conclusion}
\label{sec:conclusion}


In this work, we have brought batch-parallelism to Multicore OCaml by
introducing \tool, a library providing a lightweight embedding of
implicit batching and a toolkit for constructing batch-parallel
structures.
The key observation behind our library is that the inversion of
control provided by simple \code{async}/\code{await} asynchronous
programming primitives is entirely sufficient to transform
explicitly-batched interfaces into an implicitly-batched ones; we have
demonstrated that as such \tool can easily be ported to other modern
programming languages, such as Rust.
Beyond this embedding, \tool investigates a methodology for gradually
incorporating batch-parallelism into an existing codebases and
structures, presenting, amongst which, two general patterns for
decomposing the parallelisation of common search-structures, and
instantiating these for widely-used structures such as X-fast and
Y-fast tries, and van Emde Boas trees.
Our experimental results show the resulting concurrent data-structures
constructed using the \tool framework far outperform their
coarse-grained counterparts and scale more gracefully as parallelism
increases.

Inspired by this technique, in the future, we are planning to explore
the possibilities it presents for automatically synthesising efficient
parallel data structures that are correct by construction.



\begin{acks}
  We are grateful to Arthur Wendling who hinted the initial design of
  \tool in Multicore OCaml. We thank Matthew Flatt and George
  P\^{i}rlea for their feedback on earlier drafts of this paper.
  We also thank the anonymous reviewers of ICFP'23 and OOPSLA'24 for
  their constructive and insightful comments.
  This work was partially supported by a Singapore Ministry of
  Education (MoE) Tier 3 grant ``Automated Program Repair''
  MOE-MOET32021-0001.
\end{acks}

\section*{Data Availability}

The software artefact accompanying this paper is available
online~\cite{obatcher}.
The artefact contains the source code of \tool (in OCaml and Rust),
implementations of all data structures described in
\autoref{sec:examples}, and build scripts for reproducing the
evaluation results from \autoref{sec:evaluation}.

\bibliography{references}


\begin{thebibliography}{3}


\ifx \showCODEN    \undefined \def \showCODEN     #1{\unskip}     \fi
\ifx \showDOI      \undefined \def \showDOI       #1{#1}\fi
\ifx \showISBNx    \undefined \def \showISBNx     #1{\unskip}     \fi
\ifx \showISBNxiii \undefined \def \showISBNxiii  #1{\unskip}     \fi
\ifx \showISSN     \undefined \def \showISSN      #1{\unskip}     \fi
\ifx \showLCCN     \undefined \def \showLCCN      #1{\unskip}     \fi
\ifx \shownote     \undefined \def \shownote      #1{#1}          \fi
\ifx \showarticletitle \undefined \def \showarticletitle #1{#1}   \fi
\ifx \showURL      \undefined \def \showURL       {\relax}        \fi
\providecommand\bibfield[2]{#2}
\providecommand\bibinfo[2]{#2}
\providecommand\natexlab[1]{#1}
\providecommand\showeprint[2][]{arXiv:#2}

\bibitem[Agrawal et~al\mbox{.}(2014)]%
        {AgrawalFSSU14}
\bibfield{author}{\bibinfo{person}{Kunal Agrawal}, \bibinfo{person}{Jeremy~T.
  Fineman}, \bibinfo{person}{Brendan Sheridan}, \bibinfo{person}{Jim Sukha},
  {and} \bibinfo{person}{Robert Utterback}.} \bibinfo{year}{2014}\natexlab{}.
\newblock \showarticletitle{{Provably Good Scheduling for Parallel Programs
  that Use Data Structures through Implicit Batching}}. In
  \bibinfo{booktitle}{\emph{PPoPP}}. \bibinfo{publisher}{{ACM}},
  \bibinfo{pages}{389--390}.
\newblock
\urldef\tempurl%
\url{https://doi.org/10.1145/2555243.2555284}
\showDOI{\tempurl}


\bibitem[Le et~al\mbox{.}(2024)]%
        {obatcher}
\bibfield{author}{\bibinfo{person}{Callista Le}, \bibinfo{person}{Kiran
  Gopinathan}, \bibinfo{person}{Koon~Wen Lee}, \bibinfo{person}{Seth Gilbert},
  {and} \bibinfo{person}{Ilya Sergey}.} \bibinfo{year}{2024}\natexlab{}.
\newblock \bibinfo{booktitle}{\emph{\tool: Implementation, Data Structures, and
  Experiments (OOPSLA'24 Artefact)}}.
\newblock
\urldef\tempurl%
\url{https://doi.org/10.5281/zenodo.XXXXX}
\showDOI{\tempurl}


\bibitem[Pugh(1990)]%
        {pugh1990skip}
\bibfield{author}{\bibinfo{person}{William Pugh}.}
  \bibinfo{year}{1990}\natexlab{}.
\newblock \showarticletitle{{Skip Lists: A Probabilistic Alternative to
  Balanced Trees}}.
\newblock \bibinfo{journal}{\emph{Commun. ACM}} \bibinfo{volume}{33},
  \bibinfo{number}{6} (\bibinfo{year}{1990}), \bibinfo{pages}{668--676}.
\newblock
\urldef\tempurl%
\url{https://doi.org/10.1145/78973.78977}
\showDOI{\tempurl}


\end{thebibliography}



\begin{thebibliography}{54}


\ifx \showCODEN    \undefined \def \showCODEN     #1{\unskip}     \fi
\ifx \showDOI      \undefined \def \showDOI       #1{#1}\fi
\ifx \showISBNx    \undefined \def \showISBNx     #1{\unskip}     \fi
\ifx \showISBNxiii \undefined \def \showISBNxiii  #1{\unskip}     \fi
\ifx \showISSN     \undefined \def \showISSN      #1{\unskip}     \fi
\ifx \showLCCN     \undefined \def \showLCCN      #1{\unskip}     \fi
\ifx \shownote     \undefined \def \shownote      #1{#1}          \fi
\ifx \showarticletitle \undefined \def \showarticletitle #1{#1}   \fi
\ifx \showURL      \undefined \def \showURL       {\relax}        \fi
\providecommand\bibfield[2]{#2}
\providecommand\bibinfo[2]{#2}
\providecommand\natexlab[1]{#1}
\providecommand\showeprint[2][]{arXiv:#2}

\bibitem[Acar et~al\mbox{.}(2020)]%
        {AcarABDW20}
\bibfield{author}{\bibinfo{person}{Umut~A. Acar}, \bibinfo{person}{Daniel
  Anderson}, \bibinfo{person}{Guy~E. Blelloch}, \bibinfo{person}{Laxman
  Dhulipala}, {and} \bibinfo{person}{Sam Westrick}.}
  \bibinfo{year}{2020}\natexlab{}.
\newblock \showarticletitle{{Parallel Batch-Dynamic Trees via Change
  Propagation}}. In \bibinfo{booktitle}{\emph{ESA}}
  \emph{(\bibinfo{series}{LIPIcs}, Vol.~\bibinfo{volume}{173})}.
  \bibinfo{publisher}{Schloss Dagstuhl - Leibniz-Zentrum f{\"{u}}r Informatik},
  \bibinfo{pages}{2:1--2:23}.
\newblock
\urldef\tempurl%
\url{https://doi.org/10.4230/LIPIcs.ESA.2020.2}
\showDOI{\tempurl}


\bibitem[Adams(1993)]%
        {Adams1993}
\bibfield{author}{\bibinfo{person}{Stephen Adams}.}
  \bibinfo{year}{1993}\natexlab{}.
\newblock \showarticletitle{Functional Pearls Efficient sets—a balancing
  act}.
\newblock \bibinfo{journal}{\emph{Journal of Functional Programming}}
  \bibinfo{volume}{3}, \bibinfo{number}{4} (\bibinfo{year}{1993}),
  \bibinfo{pages}{553–561}.
\newblock
\urldef\tempurl%
\url{https://doi.org/10.1017/S0956796800000885}
\showDOI{\tempurl}


\bibitem[Adelson-Velsky and Landis(1962)]%
        {avl}
\bibfield{author}{\bibinfo{person}{Georgy Adelson-Velsky} {and}
  \bibinfo{person}{Evgenii Landis}.} \bibinfo{year}{1962}\natexlab{}.
\newblock \showarticletitle{An algorithm for the organization of information}.
\newblock \bibinfo{journal}{\emph{Proc. of the USSR Academy of Sciences}}
  \bibinfo{volume}{145} (\bibinfo{year}{1962}), \bibinfo{pages}{263–266}.
\newblock
\newblock
\shownote{In Russian, English translation by Myron J. Ricci in Soviet Doklady,
  3:1259-1263, 1962}.


\bibitem[Agrawal et~al\mbox{.}(2014)]%
        {AgrawalFSSU14}
\bibfield{author}{\bibinfo{person}{Kunal Agrawal}, \bibinfo{person}{Jeremy~T.
  Fineman}, \bibinfo{person}{Brendan Sheridan}, \bibinfo{person}{Jim Sukha},
  {and} \bibinfo{person}{Robert Utterback}.} \bibinfo{year}{2014}\natexlab{}.
\newblock \showarticletitle{{Provably Good Scheduling for Parallel Programs
  that Use Data Structures through Implicit Batching}}. In
  \bibinfo{booktitle}{\emph{PPoPP}}. \bibinfo{publisher}{{ACM}},
  \bibinfo{pages}{389--390}.
\newblock
\urldef\tempurl%
\url{https://doi.org/10.1145/2555243.2555284}
\showDOI{\tempurl}


\bibitem[Akhremtsev and Sanders(2016)]%
        {Akhremtsev2016}
\bibfield{author}{\bibinfo{person}{Yaroslav Akhremtsev} {and}
  \bibinfo{person}{Peter Sanders}.} \bibinfo{year}{2016}\natexlab{}.
\newblock \showarticletitle{Fast Parallel Operations on Search Trees}. In
  \bibinfo{booktitle}{\emph{2016 IEEE 23rd International Conference on High
  Performance Computing (HiPC)}}. \bibinfo{pages}{291--300}.
\newblock
\urldef\tempurl%
\url{https://doi.org/10.1109/HiPC.2016.042}
\showDOI{\tempurl}


\bibitem[Bayer(1972)]%
        {Bayer72}
\bibfield{author}{\bibinfo{person}{Rudolf Bayer}.}
  \bibinfo{year}{1972}\natexlab{}.
\newblock \showarticletitle{Symmetric Binary B-Trees: Data Structure and
  Maintenance Algorithms}.
\newblock \bibinfo{journal}{\emph{Acta Informatica}}  \bibinfo{volume}{1}
  (\bibinfo{year}{1972}), \bibinfo{pages}{290--306}.
\newblock
\urldef\tempurl%
\url{https://doi.org/10.1007/BF00289509}
\showDOI{\tempurl}


\bibitem[Bayer and McCreight(1972)]%
        {bayer1972symmetric}
\bibfield{author}{\bibinfo{person}{Rudolf Bayer} {and}
  \bibinfo{person}{Edward~M McCreight}.} \bibinfo{year}{1972}\natexlab{}.
\newblock \showarticletitle{{Symmetric Binary B-trees: Data Structure and
  Maintenance Algorithms}}. In \bibinfo{booktitle}{\emph{Acta informatica}},
  Vol.~\bibinfo{volume}{1}. Springer, \bibinfo{pages}{290--306}.
\newblock
\urldef\tempurl%
\url{https://doi.org/10.1007/BF00289509}
\showDOI{\tempurl}


\bibitem[Bayer and Schkolnick(1977)]%
        {Bayer77}
\bibfield{author}{\bibinfo{person}{Rudolf Bayer} {and} \bibinfo{person}{Mario
  Schkolnick}.} \bibinfo{year}{1977}\natexlab{}.
\newblock \showarticletitle{Concurrency of Operations on B-Trees}.
\newblock \bibinfo{journal}{\emph{Acta Informatica}}  \bibinfo{volume}{9}
  (\bibinfo{year}{1977}), \bibinfo{pages}{1--21}.
\newblock
\urldef\tempurl%
\url{https://doi.org/10.1007/BF00263762}
\showDOI{\tempurl}


\bibitem[Blelloch et~al\mbox{.}(2016)]%
        {BlellochFS16}
\bibfield{author}{\bibinfo{person}{Guy~E. Blelloch}, \bibinfo{person}{Daniel
  Ferizovic}, {and} \bibinfo{person}{Yihan Sun}.}
  \bibinfo{year}{2016}\natexlab{}.
\newblock \showarticletitle{Just Join for Parallel Ordered Sets}. In
  \bibinfo{booktitle}{\emph{SPAA}}. \bibinfo{publisher}{{ACM}},
  \bibinfo{pages}{253--264}.
\newblock
\urldef\tempurl%
\url{https://doi.org/10.1145/2935764.2935768}
\showDOI{\tempurl}


\bibitem[Blelloch et~al\mbox{.}(2022)]%
        {Blelloch22}
\bibfield{author}{\bibinfo{person}{Guy~E. Blelloch}, \bibinfo{person}{Daniel
  Ferizovic}, {and} \bibinfo{person}{Yihan Sun}.}
  \bibinfo{year}{2022}\natexlab{}.
\newblock \showarticletitle{Joinable Parallel Balanced Binary Trees}.
\newblock \bibinfo{journal}{\emph{{ACM} Trans. Parallel Comput.}}
  \bibinfo{volume}{9}, \bibinfo{number}{2} (\bibinfo{year}{2022}),
  \bibinfo{pages}{7:1--7:41}.
\newblock
\urldef\tempurl%
\url{https://doi.org/10.1145/3512769}
\showDOI{\tempurl}


\bibitem[Blelloch and Reid{-}Miller(1998)]%
        {Blelloch98}
\bibfield{author}{\bibinfo{person}{Guy~E. Blelloch} {and}
  \bibinfo{person}{Margaret Reid{-}Miller}.} \bibinfo{year}{1998}\natexlab{}.
\newblock \showarticletitle{Fast Set Operations Using Treaps}. In
  \bibinfo{booktitle}{\emph{SPAA}}. \bibinfo{publisher}{{ACM}},
  \bibinfo{pages}{16--26}.
\newblock
\urldef\tempurl%
\url{https://doi.org/10.1145/277651.277660}
\showDOI{\tempurl}


\bibitem[Braginsky and Petrank(2012)]%
        {Braginsky12}
\bibfield{author}{\bibinfo{person}{Anastasia Braginsky} {and}
  \bibinfo{person}{Erez Petrank}.} \bibinfo{year}{2012}\natexlab{}.
\newblock \showarticletitle{A lock-free B+tree}. In
  \bibinfo{booktitle}{\emph{SPAA}}. \bibinfo{publisher}{{ACM}},
  \bibinfo{pages}{58--67}.
\newblock
\urldef\tempurl%
\url{https://doi.org/10.1145/2312005.2312016}
\showDOI{\tempurl}


\bibitem[Brodal et~al\mbox{.}(1998)]%
        {BrodalTZ98}
\bibfield{author}{\bibinfo{person}{Gerth~St{\o}lting Brodal},
  \bibinfo{person}{Jesper~Larsson Tr{\"{a}}ff}, {and}
  \bibinfo{person}{Christos~D. Zaroliagis}.} \bibinfo{year}{1998}\natexlab{}.
\newblock \showarticletitle{A Parallel Priority Queue with Constant Time
  Operations}.
\newblock \bibinfo{journal}{\emph{J. Parallel Distributed Comput.}}
  \bibinfo{volume}{49}, \bibinfo{number}{1} (\bibinfo{year}{1998}),
  \bibinfo{pages}{4--21}.
\newblock
\urldef\tempurl%
\url{https://doi.org/10.1006/jpdc.1998.1425}
\showDOI{\tempurl}


\bibitem[Brown et~al\mbox{.}(2020)]%
        {Brown20}
\bibfield{author}{\bibinfo{person}{Trevor Brown}, \bibinfo{person}{Aleksandar
  Prokopec}, {and} \bibinfo{person}{Dan Alistarh}.}
  \bibinfo{year}{2020}\natexlab{}.
\newblock \showarticletitle{Non-blocking interpolation search trees with
  doubly-logarithmic running time}. In \bibinfo{booktitle}{\emph{PPoPP}}.
  \bibinfo{publisher}{{ACM}}, \bibinfo{pages}{276--291}.
\newblock
\urldef\tempurl%
\url{https://doi.org/10.1145/3332466.3374542}
\showDOI{\tempurl}


\bibitem[Cormen et~al\mbox{.}(2009)]%
        {clrs}
\bibfield{author}{\bibinfo{person}{Thomas~H. Cormen},
  \bibinfo{person}{Charles~E. Leiserson}, \bibinfo{person}{Ronald~L. Rivest},
  {and} \bibinfo{person}{Clifford Stein}.} \bibinfo{year}{2009}\natexlab{}.
\newblock \bibinfo{booktitle}{\emph{Introduction to Algorithms, 3rd Edition}}.
\newblock \bibinfo{publisher}{{MIT} Press}.
\newblock


\bibitem[Cruanes(2024)]%
        {ocaml-datalog}
\bibfield{author}{\bibinfo{person}{Simon Cruanes}.}
  \bibinfo{year}{2024}\natexlab{}.
\newblock \bibinfo{booktitle}{\emph{{An in-memory Datalog implementation for
  OCaml}}}.
\newblock
\urldef\tempurl%
\url{https://github.com/c-cube/datalog}
\showURL{%
\tempurl}
\newblock
\shownote{Accessed on 28 March 2024}.


\bibitem[Dhulipala et~al\mbox{.}(2020)]%
        {DhulipalaDKPSS20}
\bibfield{author}{\bibinfo{person}{Laxman Dhulipala}, \bibinfo{person}{David
  Durfee}, \bibinfo{person}{Janardhan Kulkarni}, \bibinfo{person}{Richard
  Peng}, \bibinfo{person}{Saurabh Sawlani}, {and} \bibinfo{person}{Xiaorui
  Sun}.} \bibinfo{year}{2020}\natexlab{}.
\newblock \showarticletitle{Parallel Batch-Dynamic Graphs: Algorithms and Lower
  Bounds}. In \bibinfo{booktitle}{\emph{SODA}}. \bibinfo{publisher}{{SIAM}},
  \bibinfo{pages}{1300--1319}.
\newblock
\urldef\tempurl%
\url{https://doi.org/10.1137/1.9781611975994.79}
\showDOI{\tempurl}


\bibitem[Driscoll et~al\mbox{.}(1988)]%
        {DriscollGST88}
\bibfield{author}{\bibinfo{person}{James~R. Driscoll},
  \bibinfo{person}{Harold~N. Gabow}, \bibinfo{person}{Ruth Shrairman}, {and}
  \bibinfo{person}{Robert~Endre Tarjan}.} \bibinfo{year}{1988}\natexlab{}.
\newblock \showarticletitle{{Relaxed Heaps: An Alternative to Fibonacci Heaps
  with Applications to Parallel Computation}}.
\newblock \bibinfo{journal}{\emph{Commun. {ACM}}} \bibinfo{volume}{31},
  \bibinfo{number}{11} (\bibinfo{year}{1988}), \bibinfo{pages}{1343--1354}.
\newblock
\urldef\tempurl%
\url{https://doi.org/10.1145/50087.50096}
\showDOI{\tempurl}


\bibitem[Ellis(1980)]%
        {Ellis}
\bibfield{author}{\bibinfo{person}{Carla~Schlatter Ellis}.}
  \bibinfo{year}{1980}\natexlab{}.
\newblock \showarticletitle{Concurrent Search and Insertion in {AVL} Trees}.
\newblock \bibinfo{journal}{\emph{{IEEE} Trans. Computers}}
  \bibinfo{volume}{29}, \bibinfo{number}{9} (\bibinfo{year}{1980}),
  \bibinfo{pages}{811--817}.
\newblock
\urldef\tempurl%
\url{https://doi.org/10.1109/TC.1980.1675680}
\showDOI{\tempurl}


\bibitem[Fatourou and Kallimanis(2011)]%
        {FatourouK11}
\bibfield{author}{\bibinfo{person}{Panagiota Fatourou} {and}
  \bibinfo{person}{Nikolaos~D. Kallimanis}.} \bibinfo{year}{2011}\natexlab{}.
\newblock \showarticletitle{A highly-efficient wait-free universal
  construction}. In \bibinfo{booktitle}{\emph{SPAA}}.
  \bibinfo{publisher}{{ACM}}, \bibinfo{pages}{325--334}.
\newblock
\urldef\tempurl%
\url{https://doi.org/10.1145/1989493.1989549}
\showDOI{\tempurl}


\bibitem[Feldman et~al\mbox{.}(2020)]%
        {FeldmanKE0NRS20}
\bibfield{author}{\bibinfo{person}{Yotam M.~Y. Feldman}, \bibinfo{person}{Artem
  Khyzha}, \bibinfo{person}{Constantin Enea}, \bibinfo{person}{Adam Morrison},
  \bibinfo{person}{Aleksandar Nanevski}, \bibinfo{person}{Noam Rinetzky}, {and}
  \bibinfo{person}{Sharon Shoham}.} \bibinfo{year}{2020}\natexlab{}.
\newblock \showarticletitle{Proving highly-concurrent traversals correct}.
\newblock \bibinfo{journal}{\emph{Proc. {ACM} Program. Lang.}}
  \bibinfo{volume}{4}, \bibinfo{number}{{OOPSLA}} (\bibinfo{year}{2020}),
  \bibinfo{pages}{128:1--128:29}.
\newblock
\urldef\tempurl%
\url{https://doi.org/10.1145/3428196}
\showDOI{\tempurl}


\bibitem[Frigo et~al\mbox{.}(1998)]%
        {FrigoLR98}
\bibfield{author}{\bibinfo{person}{Matteo Frigo}, \bibinfo{person}{Charles~E.
  Leiserson}, {and} \bibinfo{person}{Keith~H. Randall}.}
  \bibinfo{year}{1998}\natexlab{}.
\newblock \showarticletitle{{The Implementation of the Cilk-5 Multithreaded
  Language}}. In \bibinfo{booktitle}{\emph{PLDI}}. \bibinfo{publisher}{{ACM}},
  \bibinfo{pages}{212--223}.
\newblock
\urldef\tempurl%
\url{https://doi.org/10.1145/277650.277725}
\showDOI{\tempurl}


\bibitem[Gilbert and Li(2020)]%
        {GilbertL20}
\bibfield{author}{\bibinfo{person}{Seth Gilbert} {and}
  \bibinfo{person}{Lawrence Er~Lu Li}.} \bibinfo{year}{2020}\natexlab{}.
\newblock \showarticletitle{{How Fast Can You Update Your MST?}}. In
  \bibinfo{booktitle}{\emph{{SPAA}}}. \bibinfo{publisher}{{ACM}},
  \bibinfo{pages}{531--533}.
\newblock
\urldef\tempurl%
\url{https://doi.org/10.1145/3350755.3400240}
\showDOI{\tempurl}


\bibitem[Gu et~al\mbox{.}(2023)]%
        {Gu23}
\bibfield{author}{\bibinfo{person}{Yan Gu}, \bibinfo{person}{Ziyang Men},
  \bibinfo{person}{Zheqi Shen}, \bibinfo{person}{Yihan Sun}, {and}
  \bibinfo{person}{Zijin Wan}.} \bibinfo{year}{2023}\natexlab{}.
\newblock \showarticletitle{Parallel Longest Increasing Subsequence and van
  Emde Boas Trees}. In \bibinfo{booktitle}{\emph{SPAA}}.
  \bibinfo{publisher}{{ACM}}, \bibinfo{pages}{327--340}.
\newblock
\urldef\tempurl%
\url{https://doi.org/10.1145/3558481.3591069}
\showDOI{\tempurl}


\bibitem[Gupta et~al\mbox{.}(2023)]%
        {GuptaDKPZK23}
\bibfield{author}{\bibinfo{person}{Vishal Gupta},
  \bibinfo{person}{Kumar~Kartikeya Dwivedi}, \bibinfo{person}{Yugesh Kothari},
  \bibinfo{person}{Yueyang Pan}, \bibinfo{person}{Diyu Zhou}, {and}
  \bibinfo{person}{Sanidhya Kashyap}.} \bibinfo{year}{2023}\natexlab{}.
\newblock \showarticletitle{Ship your Critical Section, Not Your Data: Enabling
  Transparent Delegation with {TCLOCKS}}. In \bibinfo{booktitle}{\emph{OSDI}}.
  \bibinfo{publisher}{{USENIX} Association}, \bibinfo{pages}{1--16}.
\newblock
\urldef\tempurl%
\url{https://www.usenix.org/conference/osdi23/presentation/gupta}
\showURL{%
\tempurl}


\bibitem[Hendler et~al\mbox{.}(2010)]%
        {HendlerIST10}
\bibfield{author}{\bibinfo{person}{Danny Hendler}, \bibinfo{person}{Itai
  Incze}, \bibinfo{person}{Nir Shavit}, {and} \bibinfo{person}{Moran Tzafrir}.}
  \bibinfo{year}{2010}\natexlab{}.
\newblock \showarticletitle{{Flat Combining and the Synchronization-Parallelism
  Tradeoff}}. In \bibinfo{booktitle}{\emph{{SPAA}}}.
  \bibinfo{publisher}{{ACM}}, \bibinfo{pages}{355--364}.
\newblock
\urldef\tempurl%
\url{https://doi.org/10.1145/1810479.1810540}
\showDOI{\tempurl}


\bibitem[Herlihy et~al\mbox{.}(2007)]%
        {HerlihyLLS07}
\bibfield{author}{\bibinfo{person}{Maurice Herlihy}, \bibinfo{person}{Yossi
  Lev}, \bibinfo{person}{Victor Luchangco}, {and} \bibinfo{person}{Nir
  Shavit}.} \bibinfo{year}{2007}\natexlab{}.
\newblock \showarticletitle{{A Simple Optimistic Skiplist Algorithm}}. In
  \bibinfo{booktitle}{\emph{{SIROCCO}}} \emph{(\bibinfo{series}{LNCS},
  Vol.~\bibinfo{volume}{4474})}. \bibinfo{publisher}{Springer},
  \bibinfo{pages}{124--138}.
\newblock
\urldef\tempurl%
\url{https://doi.org/10.1007/978-3-540-72951-8\_11}
\showDOI{\tempurl}


\bibitem[Herlihy and Shavit(2008)]%
        {AoMP}
\bibfield{author}{\bibinfo{person}{Maurice Herlihy} {and} \bibinfo{person}{Nir
  Shavit}.} \bibinfo{year}{2008}\natexlab{}.
\newblock \bibinfo{booktitle}{\emph{{The Art of Multiprocessor Programming}}}.
\newblock \bibinfo{publisher}{Morgan Kaufmann}.
\newblock
\showISBNx{978-0-12-370591-4}


\bibitem[Herlihy and Wing(1990)]%
        {HerlihyW90}
\bibfield{author}{\bibinfo{person}{Maurice Herlihy} {and}
  \bibinfo{person}{Jeannette~M. Wing}.} \bibinfo{year}{1990}\natexlab{}.
\newblock \showarticletitle{{Linearizability: {A} Correctness Condition for
  Concurrent Objects}}.
\newblock \bibinfo{journal}{\emph{{ACM} Trans. Program. Lang. Syst.}}
  \bibinfo{volume}{12}, \bibinfo{number}{3} (\bibinfo{year}{1990}),
  \bibinfo{pages}{463--492}.
\newblock
\urldef\tempurl%
\url{https://doi.org/10.1145/78969.78972}
\showDOI{\tempurl}


\bibitem[Italiano et~al\mbox{.}(2019)]%
        {ItalianoLMP19}
\bibfield{author}{\bibinfo{person}{Giuseppe~F. Italiano},
  \bibinfo{person}{Silvio Lattanzi}, \bibinfo{person}{Vahab~S. Mirrokni}, {and}
  \bibinfo{person}{Nikos Parotsidis}.} \bibinfo{year}{2019}\natexlab{}.
\newblock \showarticletitle{{Dynamic Algorithms for the Massively Parallel
  Computation Model}}. In \bibinfo{booktitle}{\emph{SPAA}}.
  \bibinfo{publisher}{{ACM}}, \bibinfo{pages}{49--58}.
\newblock
\urldef\tempurl%
\url{https://doi.org/10.1145/3323165.3323202}
\showDOI{\tempurl}


\bibitem[Khalaji et~al\mbox{.}(2024)]%
        {Khalaji0DA24}
\bibfield{author}{\bibinfo{person}{Mohammad Khalaji}, \bibinfo{person}{Trevor
  Brown}, \bibinfo{person}{Khuzaima Daudjee}, {and} \bibinfo{person}{Vitaly
  Aksenov}.} \bibinfo{year}{2024}\natexlab{}.
\newblock \showarticletitle{Practical Hardware Transactional vEB Trees}. In
  \bibinfo{booktitle}{\emph{PPoPP}}. \bibinfo{publisher}{{ACM}},
  \bibinfo{pages}{215--228}.
\newblock
\urldef\tempurl%
\url{https://doi.org/10.1145/3627535.3638504}
\showDOI{\tempurl}


\bibitem[Le et~al\mbox{.}(2024)]%
        {obatcher}
\bibfield{author}{\bibinfo{person}{Callista Le}, \bibinfo{person}{Kiran
  Gopinathan}, \bibinfo{person}{Koon~Wen Lee}, \bibinfo{person}{Seth Gilbert},
  {and} \bibinfo{person}{Ilya Sergey}.} \bibinfo{year}{2024}\natexlab{}.
\newblock \bibinfo{booktitle}{\emph{\tool: Implementation, Data Structures, and
  Experiments (OOPSLA'24 Artefact)}}.
\newblock
\urldef\tempurl%
\url{https://doi.org/10.5281/zenodo.12604575}
\showDOI{\tempurl}


\bibitem[Lozi et~al\mbox{.}(2016)]%
        {LoziD0LM16}
\bibfield{author}{\bibinfo{person}{Jean{-}Pierre Lozi},
  \bibinfo{person}{Florian David}, \bibinfo{person}{Ga{\"{e}}l Thomas},
  \bibinfo{person}{Julia Lawall}, {and} \bibinfo{person}{Gilles Muller}.}
  \bibinfo{year}{2016}\natexlab{}.
\newblock \showarticletitle{Fast and Portable Locking for Multicore
  Architectures}.
\newblock \bibinfo{journal}{\emph{{ACM} Trans. Comput. Syst.}}
  \bibinfo{volume}{33}, \bibinfo{number}{4} (\bibinfo{year}{2016}),
  \bibinfo{pages}{13:1--13:62}.
\newblock
\urldef\tempurl%
\url{https://dl.acm.org/citation.cfm?id=2845079}
\showURL{%
\tempurl}


\bibitem[Meyer et~al\mbox{.}(2022)]%
        {MeyerWW22}
\bibfield{author}{\bibinfo{person}{Roland Meyer}, \bibinfo{person}{Thomas
  Wies}, {and} \bibinfo{person}{Sebastian Wolff}.}
  \bibinfo{year}{2022}\natexlab{}.
\newblock \showarticletitle{A concurrent program logic with a future and
  history}.
\newblock \bibinfo{journal}{\emph{Proc. {ACM} Program. Lang.}}
  \bibinfo{volume}{6}, \bibinfo{number}{{OOPSLA2}} (\bibinfo{year}{2022}),
  \bibinfo{pages}{1378--1407}.
\newblock
\urldef\tempurl%
\url{https://doi.org/10.1145/3563337}
\showDOI{\tempurl}


\bibitem[Mulder et~al\mbox{.}(2022)]%
        {MulderKG22}
\bibfield{author}{\bibinfo{person}{Ike Mulder}, \bibinfo{person}{Robbert
  Krebbers}, {and} \bibinfo{person}{Herman Geuvers}.}
  \bibinfo{year}{2022}\natexlab{}.
\newblock \showarticletitle{{Diaframe: automated verification of fine-grained
  concurrent programs in Iris}}. In \bibinfo{booktitle}{\emph{PLDI}}.
  \bibinfo{publisher}{{ACM}}, \bibinfo{pages}{809--824}.
\newblock
\urldef\tempurl%
\url{https://doi.org/10.1145/3519939.3523432}
\showDOI{\tempurl}


\bibitem[Oshman and Shavit(2013)]%
        {Oshman13}
\bibfield{author}{\bibinfo{person}{Rotem Oshman} {and} \bibinfo{person}{Nir
  Shavit}.} \bibinfo{year}{2013}\natexlab{}.
\newblock \showarticletitle{The SkipTrie: low-depth concurrent search without
  rebalancing}. In \bibinfo{booktitle}{\emph{PODC}},
  \bibfield{editor}{\bibinfo{person}{Panagiota Fatourou} {and}
  \bibinfo{person}{Gadi Taubenfeld}} (Eds.). \bibinfo{publisher}{{ACM}},
  \bibinfo{pages}{23--32}.
\newblock
\urldef\tempurl%
\url{https://doi.org/10.1145/2484239.2484270}
\showDOI{\tempurl}


\bibitem[Oyama et~al\mbox{.}(2000)]%
        {oyama1999executing}
\bibfield{author}{\bibinfo{person}{Yoshihiro Oyama}, \bibinfo{person}{Kenjiro
  Taura}, {and} \bibinfo{person}{Akinori Yonezawa}.}
  \bibinfo{year}{2000}\natexlab{}.
\newblock \showarticletitle{Executing Parallel Programs with Synchronization
  Bottlenecks Efficiently}. In \bibinfo{booktitle}{\emph{International Workshop
  on Parallel and Distributed Computing for Symbolic and Irregular Applications
  (PDSIA '99)}}. \bibinfo{publisher}{World Scientific},
  \bibinfo{pages}{182--204}.
\newblock


\bibitem[Paige and Kruskal(1985)]%
        {PaigeK85}
\bibfield{author}{\bibinfo{person}{Richard~C. Paige} {and}
  \bibinfo{person}{Clyde~P. Kruskal}.} \bibinfo{year}{1985}\natexlab{}.
\newblock \showarticletitle{{Parallel Algorithms for Shortest Path Problems}}.
  In \bibinfo{booktitle}{\emph{ICPP}}. \bibinfo{publisher}{{IEEE} Computer
  Society Press}, \bibinfo{pages}{14--20}.
\newblock


\bibitem[{Peyton Jones} et~al\mbox{.}(2008)]%
        {JonesLKC08}
\bibfield{author}{\bibinfo{person}{Simon~L. {Peyton Jones}},
  \bibinfo{person}{Roman Leshchinskiy}, \bibinfo{person}{Gabriele Keller},
  {and} \bibinfo{person}{Manuel M.~T. Chakravarty}.}
  \bibinfo{year}{2008}\natexlab{}.
\newblock \showarticletitle{{Harnessing the Multicores: Nested Data Parallelism
  in Haskell}}. In \bibinfo{booktitle}{\emph{{FSTTCS}}}
  \emph{(\bibinfo{series}{LIPIcs}, Vol.~\bibinfo{volume}{2})}.
  \bibinfo{publisher}{Schloss Dagstuhl - Leibniz-Zentrum f{\"{u}}r Informatik},
  \bibinfo{pages}{383--414}.
\newblock
\urldef\tempurl%
\url{https://doi.org/10.4230/LIPIcs.FSTTCS.2008.1769}
\showDOI{\tempurl}


\bibitem[Prokopec et~al\mbox{.}(2011)]%
        {ProkopecBRO11}
\bibfield{author}{\bibinfo{person}{Aleksandar Prokopec}, \bibinfo{person}{Phil
  Bagwell}, \bibinfo{person}{Tiark Rompf}, {and} \bibinfo{person}{Martin
  Odersky}.} \bibinfo{year}{2011}\natexlab{}.
\newblock \showarticletitle{{A Generic Parallel Collection Framework}}. In
  \bibinfo{booktitle}{\emph{Euro-Par}} \emph{(\bibinfo{series}{LNCS},
  Vol.~\bibinfo{volume}{6853})}. \bibinfo{publisher}{Springer},
  \bibinfo{pages}{136--147}.
\newblock
\urldef\tempurl%
\url{https://doi.org/10.1007/978-3-642-23397-5\_14}
\showDOI{\tempurl}


\bibitem[Pugh(1990)]%
        {pugh1990skip}
\bibfield{author}{\bibinfo{person}{William Pugh}.}
  \bibinfo{year}{1990}\natexlab{}.
\newblock \showarticletitle{{Skip Lists: A Probabilistic Alternative to
  Balanced Trees}}.
\newblock \bibinfo{journal}{\emph{Commun. ACM}} \bibinfo{volume}{33},
  \bibinfo{number}{6} (\bibinfo{year}{1990}), \bibinfo{pages}{668--676}.
\newblock
\urldef\tempurl%
\url{https://doi.org/10.1145/78973.78977}
\showDOI{\tempurl}


\bibitem[Roghanchi et~al\mbox{.}(2017)]%
        {RoghanchiEB17}
\bibfield{author}{\bibinfo{person}{Sepideh Roghanchi}, \bibinfo{person}{Jakob
  Eriksson}, {and} \bibinfo{person}{Nilanjana Basu}.}
  \bibinfo{year}{2017}\natexlab{}.
\newblock \showarticletitle{ffwd: delegation is (much) faster than you think}.
  In \bibinfo{booktitle}{\emph{SOSP}}. \bibinfo{publisher}{{ACM}},
  \bibinfo{pages}{342--358}.
\newblock
\urldef\tempurl%
\url{https://doi.org/10.1145/3132747.3132771}
\showDOI{\tempurl}


\bibitem[Sanders et~al\mbox{.}(2019)]%
        {Sanders2019}
\bibfield{author}{\bibinfo{person}{Peter Sanders}, \bibinfo{person}{Kurt
  Mehlhorn}, \bibinfo{person}{Martin Dietzfelbinger}, {and}
  \bibinfo{person}{Roman Dementiev}.} \bibinfo{year}{2019}\natexlab{}.
\newblock \bibinfo{booktitle}{\emph{Sorted Sequences}}.
\newblock \bibinfo{publisher}{Springer International Publishing},
  \bibinfo{address}{Cham}, \bibinfo{pages}{233--258}.
\newblock
\showISBNx{978-3-030-25209-0}
\urldef\tempurl%
\url{https://doi.org/10.1007/978-3-030-25209-0_7}
\showDOI{\tempurl}


\bibitem[Seidel and Aragon(1996)]%
        {SeidelA96}
\bibfield{author}{\bibinfo{person}{Raimund Seidel} {and}
  \bibinfo{person}{Cecilia~R. Aragon}.} \bibinfo{year}{1996}\natexlab{}.
\newblock \showarticletitle{Randomized Search Trees}.
\newblock \bibinfo{journal}{\emph{Algorithmica}} \bibinfo{volume}{16},
  \bibinfo{number}{4/5} (\bibinfo{year}{1996}), \bibinfo{pages}{464--497}.
\newblock
\urldef\tempurl%
\url{https://doi.org/10.1007/BF01940876}
\showDOI{\tempurl}


\bibitem[Sergey et~al\mbox{.}(2015)]%
        {SergeyNB15}
\bibfield{author}{\bibinfo{person}{Ilya Sergey}, \bibinfo{person}{Aleksandar
  Nanevski}, {and} \bibinfo{person}{Anindya Banerjee}.}
  \bibinfo{year}{2015}\natexlab{}.
\newblock \showarticletitle{{Mechanized Verification of Fine-Grained Concurrent
  Programs}}. In \bibinfo{booktitle}{\emph{PLDI}}. \bibinfo{publisher}{{ACM}},
  \bibinfo{pages}{77--87}.
\newblock
\urldef\tempurl%
\url{https://doi.org/10.1145/2737924.2737964}
\showDOI{\tempurl}


\bibitem[Sivaramakrishnan et~al\mbox{.}(2021)]%
        {Sivaramakrishnan21}
\bibfield{author}{\bibinfo{person}{K.~C. Sivaramakrishnan},
  \bibinfo{person}{Stephen Dolan}, \bibinfo{person}{Leo White},
  \bibinfo{person}{Tom Kelly}, \bibinfo{person}{Sadiq Jaffer}, {and}
  \bibinfo{person}{Anil Madhavapeddy}.} \bibinfo{year}{2021}\natexlab{}.
\newblock \showarticletitle{Retrofitting effect handlers onto OCaml}. In
  \bibinfo{booktitle}{\emph{PLDI}}. \bibinfo{publisher}{{ACM}},
  \bibinfo{pages}{206--221}.
\newblock
\urldef\tempurl%
\url{https://doi.org/10.1145/3453483.3454039}
\showDOI{\tempurl}


\bibitem[Srivastava and Brown(2022)]%
        {Srivastava22}
\bibfield{author}{\bibinfo{person}{Anubhav Srivastava} {and}
  \bibinfo{person}{Trevor Brown}.} \bibinfo{year}{2022}\natexlab{}.
\newblock \showarticletitle{Elimination (a, b)-trees with fast, durable
  updates}. In \bibinfo{booktitle}{\emph{PPoPP}}. \bibinfo{publisher}{{ACM}},
  \bibinfo{pages}{416--430}.
\newblock
\urldef\tempurl%
\url{https://doi.org/10.1145/3503221.3508441}
\showDOI{\tempurl}


\bibitem[Treiber(1986)]%
        {Treiber:TR}
\bibfield{author}{\bibinfo{person}{R.~Kent Treiber}.}
  \bibinfo{year}{1986}\natexlab{}.
\newblock \bibinfo{booktitle}{\emph{Systems programming: coping with
  parallelism}}.
\newblock \bibinfo{type}{{T}echnical {R}eport} RJ 5118.
  \bibinfo{institution}{IBM Almaden Research Center}.
\newblock


\bibitem[Tseng et~al\mbox{.}(2019)]%
        {TsengDB19}
\bibfield{author}{\bibinfo{person}{Thomas Tseng}, \bibinfo{person}{Laxman
  Dhulipala}, {and} \bibinfo{person}{Guy Blelloch}.}
  \bibinfo{year}{2019}\natexlab{}.
\newblock \showarticletitle{Batch-Parallel Euler Tour Trees}. In
  \bibinfo{booktitle}{\emph{Proceedings of the Twenty-First Workshop on
  Algorithm Engineering and Experiments ({ALENEX})}}.
  \bibinfo{publisher}{{SIAM}}, \bibinfo{pages}{92--106}.
\newblock
\urldef\tempurl%
\url{https://doi.org/10.1137/1.9781611975499.8}
\showDOI{\tempurl}


\bibitem[Tseng et~al\mbox{.}(2022)]%
        {TsengDS22}
\bibfield{author}{\bibinfo{person}{Tom Tseng}, \bibinfo{person}{Laxman
  Dhulipala}, {and} \bibinfo{person}{Julian Shun}.}
  \bibinfo{year}{2022}\natexlab{}.
\newblock \showarticletitle{Parallel Batch-Dynamic Minimum Spanning Forest and
  the Efficiency of Dynamic Agglomerative Graph Clustering}. In
  \bibinfo{booktitle}{\emph{SPAA}}. \bibinfo{publisher}{ACM},
  \bibinfo{pages}{233–245}.
\newblock
\urldef\tempurl%
\url{https://doi.org/10.1145/3490148.3538584}
\showDOI{\tempurl}


\bibitem[Vafeiadis(2008)]%
        {Vafeiadis08}
\bibfield{author}{\bibinfo{person}{Viktor Vafeiadis}.}
  \bibinfo{year}{2008}\natexlab{}.
\newblock \emph{\bibinfo{title}{Modular fine-grained concurrency
  verification}}.
\newblock \bibinfo{thesistype}{Ph.\,D. Dissertation}.
  \bibinfo{school}{University of Cambridge}.
\newblock


\bibitem[van Emde~Boas(1975)]%
        {VanEmdeBoas1977}
\bibfield{author}{\bibinfo{person}{Peter van Emde~Boas}.}
  \bibinfo{year}{1975}\natexlab{}.
\newblock \showarticletitle{{Preserving Order in a Forest in less than
  Logarithmic Time}}. In \bibinfo{booktitle}{\emph{FOCS}}.
  \bibinfo{publisher}{{IEEE} Computer Society}, \bibinfo{pages}{75--84}.
\newblock
\urldef\tempurl%
\url{https://doi.org/10.1109/SFCS.1975.26}
\showDOI{\tempurl}


\bibitem[Wang et~al\mbox{.}(2021)]%
        {WangYGS21}
\bibfield{author}{\bibinfo{person}{Yiqiu Wang}, \bibinfo{person}{Shangdi Yu},
  \bibinfo{person}{Yan Gu}, {and} \bibinfo{person}{Julian Shun}.}
  \bibinfo{year}{2021}\natexlab{}.
\newblock \showarticletitle{A{ Parallel Batch-Dynamic Data Structure for the
  Closest Pair Problem}}. In \bibinfo{booktitle}{\emph{37th International
  Symposium on Computational Geometry (SoCG 2021)}}
  \emph{(\bibinfo{series}{LIPIcs}, Vol.~\bibinfo{volume}{189})}.
  \bibinfo{publisher}{Schloss Dagstuhl - Leibniz-Zentrum f{\"{u}}r Informatik},
  \bibinfo{pages}{60:1--60:16}.
\newblock
\urldef\tempurl%
\url{https://doi.org/10.4230/LIPIcs.SoCG.2021.60}
\showDOI{\tempurl}


\bibitem[Willard(1983)]%
        {Willard1983}
\bibfield{author}{\bibinfo{person}{Dan~E. Willard}.}
  \bibinfo{year}{1983}\natexlab{}.
\newblock \showarticletitle{Log-logarithmic worst-case range queries are
  possible in space O(N)}.
\newblock \bibinfo{journal}{\emph{Inform. Process. Lett.}}
  \bibinfo{volume}{17}, \bibinfo{number}{2} (\bibinfo{year}{1983}),
  \bibinfo{pages}{81--84}.
\newblock
\showISSN{0020-0190}
\urldef\tempurl%
\url{https://doi.org/10.1016/0020-0190(83)90075-3}
\showDOI{\tempurl}


\end{thebibliography}

\appendix
\section{Batch-Parallel Skip List}
\label{sec:skiplist}

\begin{wrapfigure}[8]{r}{0.45\linewidth}
\vspace{-15pt}
\setlength{\abovecaptionskip}{-5pt}
\includegraphics[width=1.0\linewidth]{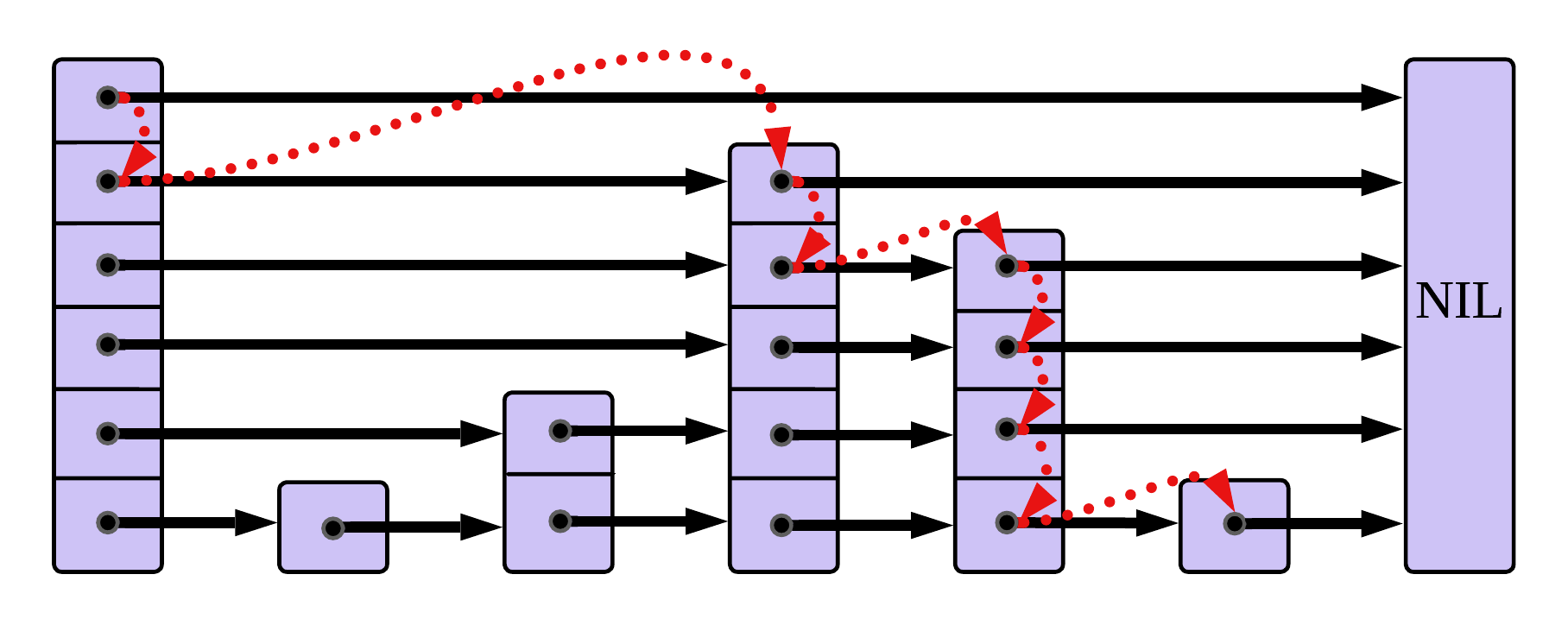}
\label{fig:skiplist-search}
\caption{Searching in a skip list}
\label{fig:slsearch}
\end{wrapfigure}
Skip lists~\cite{pugh1990skip} are collections of ordered lists that
allow for logarithmic-time inserts, search, and deletion.
%
%
%
Unlike vanilla sorted lists that only permit walking through the
entire list to reach some node, the skip list algorithm finds a faster
path by \emph{skipping} nodes.
This is possible thanks to logically grouping pointers in skip lists
into ``levels'' whereby each higher level is a sub-list of the level
right below it. 
%
%
To find a node within the skip list, the search operation traversing
it at the highest level, progressively checking if the key is smaller
than the next node.
It keeps advancing while the key is smaller, and once it becomes
larger than the next node's value, the search procedure moves the
traversal to a more densely populated level below, starting from the
last visited node with a smaller key (\cf~\autoref{fig:slsearch}).
It proceeds by moving down the levels until it either locates the
sought node or reaches the lowest level. At that point, it either
locates the node by the last traversal or discovers that the desired
key is not present in the list.
The insert operation works in the same way as the search except that
it must keep track of all previous nodes that will have their forward
pointers adjusted to the inserted node. Therefore, a node with $n$
levels where $n$ is randomly determined, will have to equivalently
adjust $n$ pointers.


\paragraph{Batch-parallel implementation}
Our batch-parallel implementation adopts the skip list design for
Cilk-5 from the work by~\citet{AgrawalFSSU14} on implicit batching. 
%
%
Since search operations are safe to run in parallel with each other,
we schedule all such operations in a batch to execute and terminate
before we start executing insertions.
The batch-parallel implementation of insertion for skip-lists runs in
thee steps:
(1) build an intermediary skip list from the batch;
(2) perform a parallel search to find the actual slots in the original
lists where the new nodes should go, and
(3) merge the intermediary list into the original list.
Below, we explain those steps in detail.

\begin{figure}[t]
\begin{subfigure}{0.49\textwidth}
\begin{minipage}{0.98\textwidth}
\includegraphics[width=1.0\linewidth]{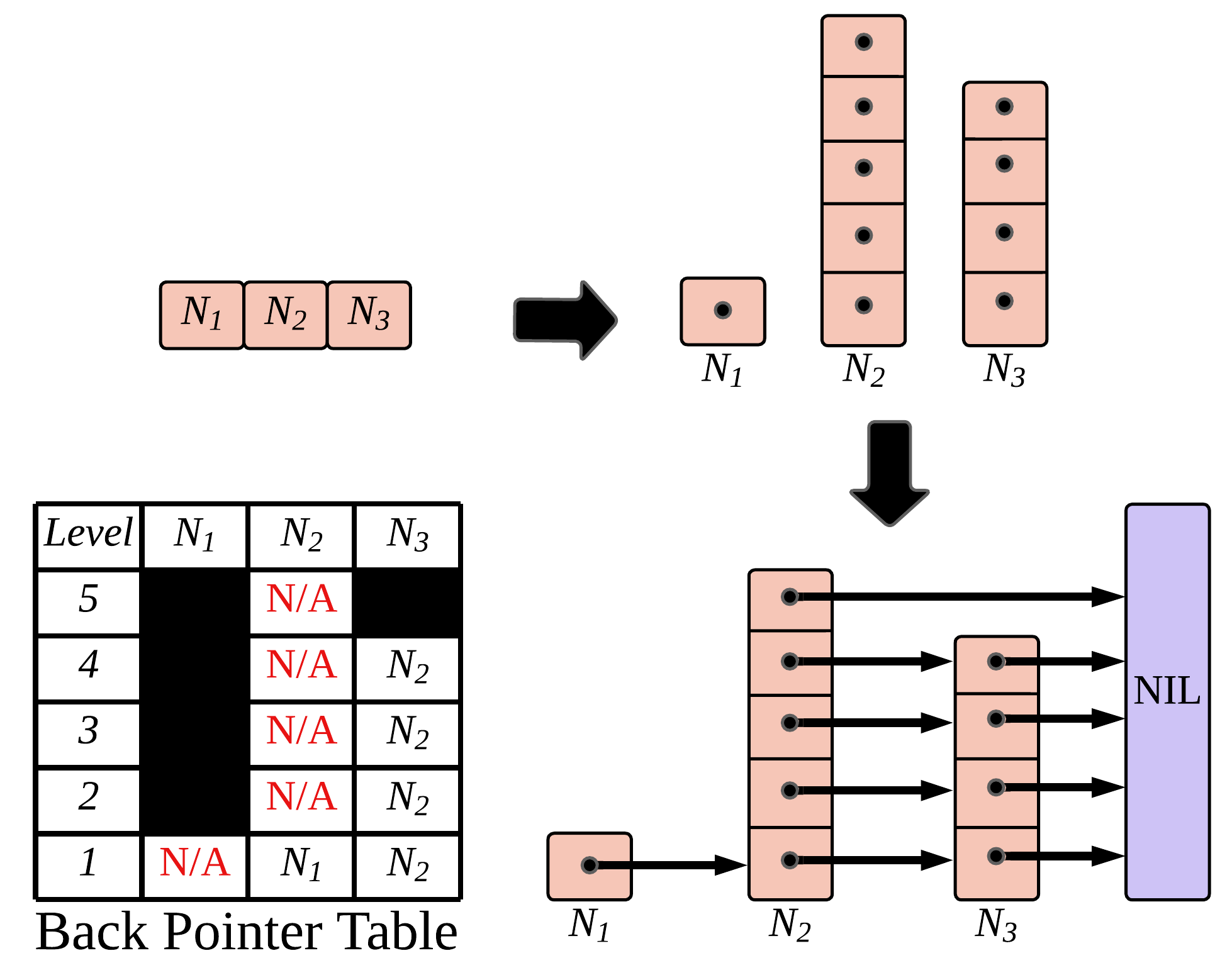}
\end{minipage}
\caption{Generating intermediary skip list from a batch}
\label{fig:intermediary}
\end{subfigure}
\begin{subfigure}{0.49\textwidth}
\begin{minipage}{1.0\textwidth}
\includegraphics[width=1.0\linewidth]{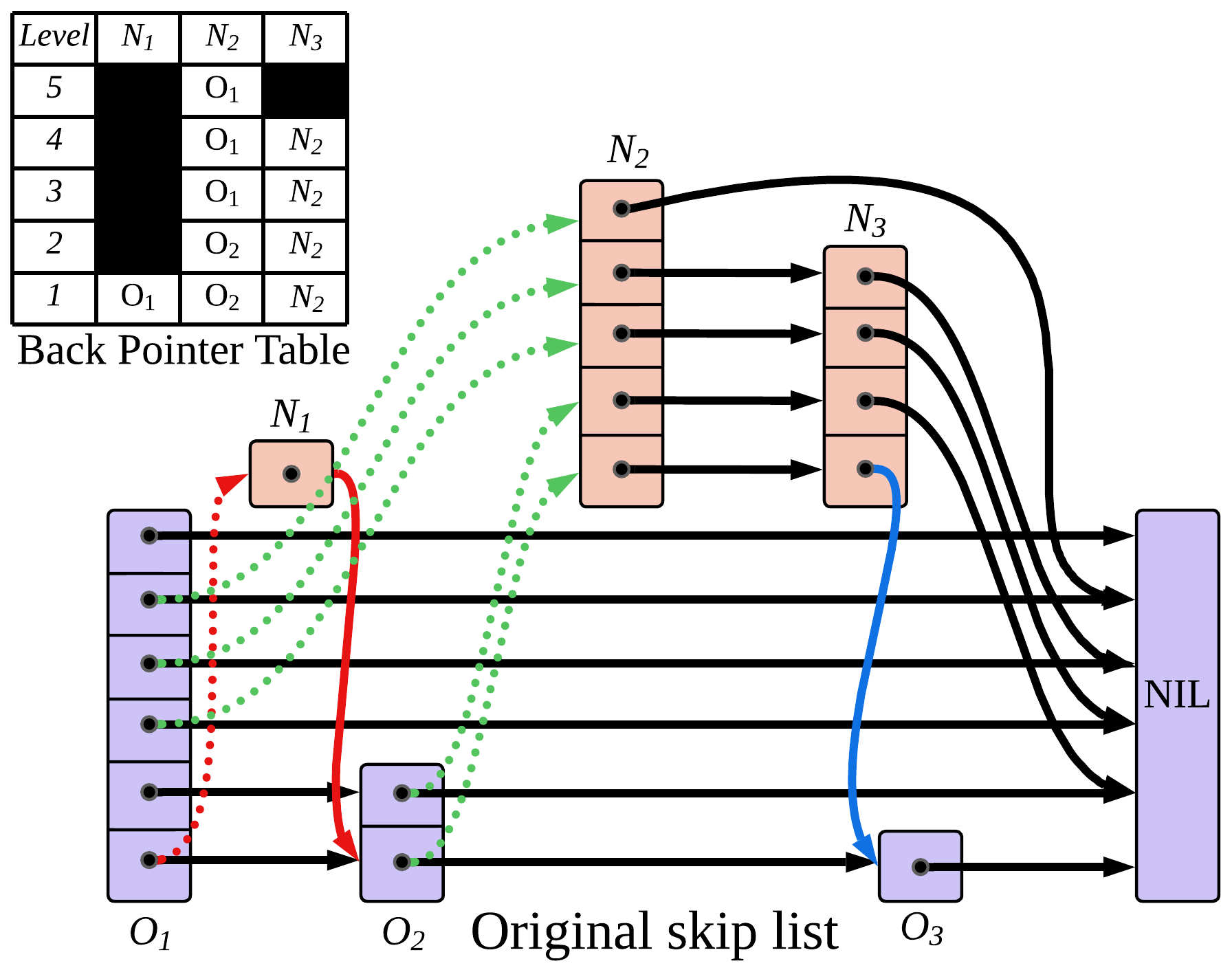}
\end{minipage}
\caption{Parallel search and update of an intermediary list}
\label{fig:skiplist-par-search}
\end{subfigure}
\caption{Two first stages of batch-parallel insertion into a skip list}
\label{fig:slinsert}
\end{figure}

\vspace{3pt}
\emph{(1) Intermediary skip list creation.} 
We begin preparing the batch by performing a parallel sort, dropping
any duplicates in the process (skip lists implement sets).
Next we pre-allocate nodes corresponding to each value to be inserted.
The level of the allocated node is determined at random.
Lastly, we construct the internal links between nodes and, for each
node, store its \emph{back pointers} (\ie, the nodes that precede it
in the list) for all the levels it's present at.
An example is shown in \autoref{fig:intermediary}: given node $N_1$
that precedes $N_2$, $N_1$ is the back pointer for $N_2$.
%

\vspace{3pt}
\emph{(2) Parallel search and update.}
The next stage prepares the intermediary skip list for being merged
into the original one.
The goal of our parallel search is two-fold: 
(a) to find the back pointers (preceding nodes) for those nodes in the
intermediary list that have none and (b) readjust the links in the
intermediate list that should point to nodes from the original list.

To implement this logic, we run a non-mutating parallel search that
looks for an appropriate slot \emph{for each new node} in the original
list whilst additionally keeping track of the back pointers for all
nodes from the original list.
As a byproduct of this search, we also obtain forward pointers for
each new node based on its discovered location in the original list.
In the case where we find a new forward pointer at a corresponding level in
the original list, we check if the target node value in the
intermediate list is \emph{smaller} than the target value in the
original list.
If this the case, we know that after merging two lists these links
will be preserved, hence no action is required; if not, we perform the
necessary remapping from a target node in an intermediary list into
the original list.

Consider the example depicted in \autoref{fig:slinsert}. 
In our illustration, the lines in red, blue and green represent
different threads that are running in parallel to perform the search.
The solid lines are pointers that are installed fully and the dotted
lines are pointers that are going to be installed. 
Notice that the new node $N_1$ to be added has no back pointer
(\autoref{fig:intermediary}).
Our parallel search-and-update procedure identifies $O_1$ in the
original list is the back pointer for $N_1$ and updates the table
correspondingly.
Regarding the forward pointers of $N_1$, notice the link
$N_1 \rightarrow N_2$ in the intermediary list should be broken to
incorporate the node $O_2$ from the original list, as shown in
\autoref{fig:skiplist-par-search}.

It crucial that \emph{no modifications} are made to the original skip
list during the parallel search and update. 
If that were not the case, data races in the original list would
affect the algorithm's ability to find correct slots.
At the same time, it is safe to modify the intermediate skip list
because no memory sharing occurs between parallel threads that handle
the case of each of the new nodes.

\vspace{3pt}
\emph{(3) Merge.}
The last step is to sequentially walk over the back pointer table and
perform the linking into the original skip list.
This step can also be performed in parallel since all the correct
mappings have been found. However, since this is an inexpensive
procedure compared to searching for locations to install nodes, we
have left it sequential.

\vspace{3pt}

Even though the strategy for parallelising inserts adopted in this
case study is fairly simple and only concerns step (2), \ie, search
and update, it still allows for substantial improvements in the
concurrent throughput of the structure.


\end{document}